\documentclass[12pt,preprint]{aastex}
\usepackage{apjfonts}
\usepackage{graphicx}
\usepackage{epstopdf}
\shorttitle{Three-phase hot-core model with glycine}
\shortauthors{R. T. Garrod}

\begin{document}

\title{A three-phase chemical model of hot cores: \\
the formation of glycine}

\author{Robin T. Garrod}
\affil{Center for Radiophysics and Space Research, Cornell University, Ithaca, NY 14853-6801,
USA} \email{rgarrod@astro.cornell.edu}

\begin{abstract}

A new chemical model is presented that simulates fully-coupled gas-phase, grain-surface and bulk-ice chemistry in hot cores. Glycine (NH$_2$CH$_2$COOH), the simplest amino acid, and related molecules such as glycinal, propionic acid and propanal, are included in the chemical network. Glycine is found to form in moderate abundance within and upon dust-grain ices via three radical-addition mechanisms, with no single mechanism strongly dominant. Glycine production in the ice occurs over temperatures $\sim$40--120 K. Peak gas-phase glycine fractional abundances lie in the range $8\times 10^{-11}$--$8\times 10^{-9}$, occuring at $\sim$200 K, the evaporation temperature of glycine. A gas-phase mechanism for glycine production is tested and found insignificant, even under optimal conditions. A new spectroscopic radiative-transfer model is used, allowing the translation and comparison of the chemical-model results with observations of specific sources. Comparison with the nearby hot-core source NGC 6334 IRS1 shows excellent agreement with integrated line intensities of observed species, including methyl formate. The results for glycine are consistent with the current lack of a detection of this molecule toward other sources; the high evaporation temperature of glycine renders the emission region extremely compact. Glycine detection with ALMA is predicted to be highly plausible, for bright, nearby sources with narrow emission lines. Photodissociation of water and subsequent hydrogen-abstraction from organic molecules by OH, and NH$_2$, are crucial to the build-up of complex organic species in the ice. The inclusion of alternative branches within the network of radical-addition reactions appears important to the abundances of hot-core molecules; less favorable branching ratios may remedy the anomalously high abundance of glycolaldehyde predicted by this and previous models.

\end{abstract}


\section{Introduction}

Millimeter, sub-millimeter, and centimeter wavelength observations of star-forming sources known as ``hot cores'' reveal a range of complex organic molecules, through the identification of molecular rotational emission spectra (Herbst \& van Dishoeck 2009). Many of the molecules detected toward hot cores are highly saturated (i.e. hydrogen-rich), and a significant number also contain oxygen and/or nitrogen; examples include methanol (CH$_3$OH), formic acid (HCOOH), and the ubiquitous methyl formate (HCOOCH$_3$), as well as the recently detected aminoacetonitrile (NH$_2$CH$_2$CN, Belloche et al. 2008), which bears both amine and nitrile groups. The breadth of functional groups borne by molecules detected toward hot cores (as well as the large number of unidentified emission lines) suggests that other, as-yet undetected species may also be present, including biologically-significant molecules such as amino acids -- the building blocks of proteins.

The search for the simplest amino acid, glycine (NH$_2$CH$_2$COOH), toward interstellar sources has not so far yielded a detection (Snyder et al. 2005); recent upper limits toward Sgr B2(N), a spectrally- and chemically-rich hot-core source in the Galactic Center, are around 10$^{14}$ -- 10$^{15}$ cm$^{-2}$ for beam-averaged values obtained with the Mopra telescope (Cunningham et al. 2007) and the ATCA array (Jones et al. 2007) for glycine conformer I. Upper limits on the higher-energy conformer II from the same studies are around 10$^{13}$ -- 10$^{14}$ cm$^{-2}$. However, numerous amino acids have been found in meteorites, and glycine was recently detected in a sample taken from comet Wild 2 (Elsila et al.\ 2009); it has indeed been postulated that terrestrial amino acids may have an interstellar origin (e.g. Ehrenfreund \& Charnley 2000).

It is likely that amino acids and other biologically-relevant species will, in the coming years, provide plausible targets for searches with the ALMA telescope, which can provide much greater sensitivity and spatial resolution than was previously available. It is therefore valuable to extend existing astrochemical models to simulate the putative formation of glycine and related species in hot cores, both to make quantitiative predictions and to assess the viability of observational candidate sources.

Current models of hot-core chemistry place great emphasis on the formation of organic molecules within or upon dust-grain ice mantles, as well as in the gas phase. Garrod \& Herbst (2006, hereafter G\&H) showed that typically-observed molecules such as methyl formate, while difficult to form by gas-phase mechanisms, could be produced in appropriate abundances through grain-surface chemistry at intermediate temperatures, prior to the complete sublimation of the dust-grain ices. The radicals HCO and CH$_3$ were found to become mobile around 20 -- 40 K, reacting with other radicals associated with formaldehyde and methanol destruction, to produce formic acid, dimethyl ether and methyl formate. Cosmic ray-induced photolysis of the molecular ice mantles was found to be the dominant source of these functional-group radicals. Garrod, Widicus Weaver \& Herbst (2008, hereafter GWH) extended the network to include a much larger number of surface radical species, allowing the study of more complex molecules; methanol was found to be the key molecule for the supply of the constituent functional groups of many such species. 

Recent experimental work by {\"O}berg et al. (2009a) has provided direct evidence that a range of complex organic molecules detected in star-forming regions may be readily produced within UV-irradiated organic ices. Other studies focused specifically on amino-acid formation have elucidated a number of potential chemical pathways for glycine within laboratory ice mixtures. Experiments by Bernstein et al. (2002) and Mu{\~n}oz Caro et al. (2002) showed that UV irradiation of interstellar ice analogs containing H$_2$O, CH$_3$OH, CO, CO$_2$, NH$_3$ and/or HCN could produce significant quantities of amino acids, including glycine. Sorrell (2001) had suggested a route involving the reaction of the NH$_2$ radical with acetic acid (CH$_3$COOH), itself formed through the addition of CH$_3$ and HOCO, while Woon (2002) suggested that the addition reaction between radicals NH$_2$CH$_2$ and HOCO could explain the formation of glycine, with these species produced in the ices by successive hydrogenation of HCN and by the addition of CO and OH, respectively. Experimental work by Holtom et al. (2005) involving the electron bombardment of CO$_2$/NH$_2$CH$_3$ ices confirmed the viability of Woon's mechanism, while work by Elsila et al. (2007), extending the investigation of Bernstein et al. (2002), used isotopic labeling to trace the origins of the constituent atoms in the product molecules. Their findings indicated that Woon's mechanism was consistent with the origins of the nitrogen and central carbon in the glycine molecule, while the origin of the acidic carbon was more consistent with the formation of glycine occurring via a nitrile precursor, i.e. amino acetonitrile (NH$_2$CH$_2$CN). However, a Strecker-type formation route was found to be of only minor importance.

In order to investigate the formation of glycine within dust-grain ices under astrophysical conditions, the extension of the chemical networks of G\&H and GWH -- as well as that of Belloche et al. (2009), who treated the formation of amino acetonitrile, $n$-propyl cyanide, and ethyl formate -- is continued with the addition of glycine, its amino aldehyde glycinal, and a collection of related species such as propionic acid and propanal. The new formation mechanisms for these species follow the general treatments of previous networks, which allowed radicals on the grains, frequently produced by UV photodissociation or chemical abstraction processes, to meet via thermal diffusion and thus form more complex structures. The mechanism suggested by Woon (2002) is included in the new network, as well as several other likely radical-addition mechanisms. Furthermore, a gas-phase mechanism for glycine production is tested, by the inclusion of reactions between protonated hydroxylamine (NH$_3$OH$^+$) and acetic acid (CH$_3$COOH), followed by dissociative electronic recombination, as suggested by Blagojevic et al. (2003). All new species in the network are afforded both formation and destruction mechanisms on the grains and in the gas-phase. A more comprehensive hydrogen-abstraction reaction set is also included in the network.

To apply this chemical network to hot-core conditions, a new three-phase astrochemical model, {\em MAGICKAL}, is used, which includes a fully-active chemistry within the bulk of the ice mantle (see Section 2.1). The physico-chemical evolution of the core is traced starting from the free-fall collapse of a cloud, using the dust-temperature/visual-extinction relationship given by G\&P, through the subsequent warm-up of the dense core from 8 -- 400 K. 

In order to simulate the emission from glycine and a selection of other molecules, the results of these models are mapped to existing observationally-determined temperature and density profiles for specific sources. The equation of radiative transfer is then solved numerically, under the assumption of local thermodynamic equilibrium, along both on- and off-source lines of sight, allowing a determination of line strengths and emission radii. This emission is then convolved with telescope beams appropriate to current instruments, allowing the detectability of glycine to be assessed. The nearby hot-core source NGC 6334 IRS1 is specifically considered, due both to the availability of physical profiles and molecular observations in the literature, and to its spectroscopic suitability. Comparison is also made with the recent non-detection of hydroxylamine (NH$_2$OH) in another source (Pulliam et al. 2012), with implications for the putative formation of glycine in the gas phase.

Section 2 describes the new chemical and physical models; the results are presented in Section 3. The results of the inclusion of gas-phase formation mechanisms for glycine are described in Section 4. The spectroscopic model and results for various molecules are presented in Section 5. A broad discussion of the results is presented in Section 6; a list of general conclusions is provided in Section 7.

\section{Chemical modeling}

To model the hot-core chemistry, a new astrochemical code is employed, named {\em MAGICKAL} (Model for Astrophysical Gas and Ice Chemical Kinetics And Layering). This code uses a rate-equation/modified rate-equation approach to solve the coupled gas-phase, dust grain-surface and ice-mantle chemistry. The essential functionality of previous approaches (Garrod \& Pauly, 2011) is retained, while a number of additional features are added, as outlined below.

\subsection{Three-phase model}

The three-phase model used here is described in detail by Garrod \& Pauly (2011), and is based on the approach of Hasegawa \& Herbst (1993). In contrast to both of those methods, {\em MAGICKAL} simulates an active ice-mantle chemistry in addition to that occurring in the surface layer of the ice or on the dust-grain surface. Previous {\em two-phase} hot-core models, such as those of G\&H, GWH, and Belloche et al. (2009), have implicitly assumed a combined surface/mantle chemistry, using surface kinetics to simulate all dust-grain chemical processes, including those within the bulk ice. Although satisfactory results may be obtained through this simplification, it has become clear through experimental means (e.g. {\"O}berg et al., 2009a) that the promotion of complex-molecule formation within organic ices by UV photolysis involves sub-surface processes, even in relatively thin ices (on the order of 10 monolayers).
 
In order to simulate chemical kinetics within the ice mantles, it is here assumed that the mobility of chemical reactants in the bulk ice occurs through a barrier-mediated thermal diffusion process. Other authors have assumed a swapping mechanism to be active in laboratory ice experiments ({\"O}berg et al. 2009; Fayolle et al. 2011), which is also implicitly assumed in this model. The swapping rate is parameterized in the same way as the surface thermal hopping, i.e. 

\begin{equation}
k_{\mathrm{swap}}(i)=\nu_{0}(i) \ \exp[-E_{\mathrm{swap}}(i)/T_{d}]
\end{equation}\

\noindent where $\nu_{0}$ is the characteristic vibrational frequency of an harmonic oscillator, $T_d$ is the dust temperature, and $E_{\mathrm{swap}}$ is an energy barrier associated with the swapping of species $i$ with an adjacent water molecule, in line with the adoption of surface binding energies appropriate to amorphous water ice, following G\&H and GWH.

Ratios between the surface diffusion barrier and the desorption energy of any particular species bound to an ice or any other surface are poorly constrained, and various authors have used values in the approximate range of 0.3 -- 0.8. Garrod \& Pauly (2011) found that the production of interstellar CO$_2$ ice could be well reproduced by the adoption of ratios $E_{\mathrm{dif}}$:$E_{\mathrm{des}} \lesssim 0.4$; here a value of 0.35 is used. Thence, a value of $E_{\mathrm{swap}}$:$E_{\mathrm{des}} = 0.7$ is assumed for the swapping barrier, on the simplistic consideration that a bulk-ice molecule is bound to approximately twice the number of binding partners as a surface molecule. In analogy to the surface chemical reaction rates, the rate of reaction between mantle-species A and B is given by:

\begin{equation}
R_{\mathrm{AB}}=N_{m}(A) \ N_{m}(B) \ [k_{\mathrm{swap}}(A) + k_{\mathrm{swap}}(B)] / N_{\mathrm{M}}
\end{equation}\

\noindent where $N_{m}(i)$ is the population of species $i$ in the ice mantle, and $N_{\mathrm{M}}$ is the total population of the ice mantle, including all species.

As discussed by Garrod \& Pauly (2011), in the three-phase model, material in the surface layer is transferred to the bulk ice according to the net rate of deposition of material onto the grain from the gas phase, with a similar, alternative transfer occurring in reverse in the case of a net {\em desorption} of mantle species into the gas phase. These transfers represent the covering or exposure of ice material as the boundary of the ice-surface layer is constantly re-defined within the model, rather than the physical transport of material. However, the inclusion of bulk diffusion within the ice should allow a true exchange of chemical species between surface and mantle, through the aforementioned swapping mechanism.

This surface--mantle swapping is a pair-wise process, and therefore requires the construction of rates for each chemical species in the mantle and in the surface layer, which must collectively produce no net transfer between the two phases. This zero net rate would be assured using an explicit pair-wise coupling of every species in the surface with every species in the grain mantle; however, such an approach is found to be computationally exhaustive, and requires estimates of the pair-wise positions of the swapped species that a rate-based model cannot provide, other than through purely statistical considerations.

Instead, the swapping rates from mantle to surface are calculated first, with rates analogous to that of equation (2). The total rate in this direction is then matched with an identical total transfer rate from surface to mantle, apportioned to all the surface species according to their fractional contribution to the total surface-layer population. Thus:

\begin{equation}
R_{\mathrm{swap,m}}(i)=N_{m}(i) \ \frac{N_{S}}{N_{M}} \ k_{\mathrm{swap}}(i)
\end{equation}\

\begin{equation}
R_{\mathrm{swap,s}}(i)= \frac{N_{s}(i)}{N_{S}} \cdot \sum_{all j} R_{\mathrm{swap,m}}(j)
\end{equation}\

where $R_{\mathrm{swap,m}}(i)$ is the rate of mantle-to-surface swapping of mantle species $i$, $R_{\mathrm{swap,s}}(i)$ is the rate of surface-to-mantle swapping of surface species $i$, $N_{m}(i)$ and $N_{s}(i)$ are the mantle and surface populations, respectively, of species $i$, and $N_{M}$ and $N_{S}$ are the total mantle and surface populations. The ratio $N_{S}/N_{M}$ in equation (3) is taken as unity when it exceeds this value, representing the case where some surface species reside on the dust-grain surface itself and have no underlying mantle.

Because the model does not assign specific positions to each atom or molecule in the mantle, the rate of mantle-to-surface swapping, $R_{\mathrm{swap,m}}(i)$, is required to be representative not simply of the final diffusion step, but of the entire chain of diffusion steps by which a mantle species at arbitrary distance from the surface layer will reach the surface. The use of $k_{\mathrm{swap}}$ in eq. (3) as the rate of a typical, or average, step in this process is accurate so long as the final diffusion step from mantle to surface is relatively fast, i.e. that intra-mantle diffusion is the rate-limiting step in the entire journey from mantle to surface. Under the assumption that the ice is predominantly water ice (see above), the final surface-to-mantle step will be fast for species whose binding energies are less than or similar to that of water. For the small minority of species with significantly greater binding energies than this, for which there would thus be an associated endothermicity to the mantle--surface swap with a water molecule (see e.g. {\"O}berg et al. 2009b), the large absolute size of the swapping barrier would render the rate of mantle-to-surface diffusion negligible compared to other processes, such as the mantle-to-surface transfer associated with net surface desorption.

These rates defined in eqs. (3) and (4) are also considered in the loss and gain rates used for the modified-rate calculations of Garrod (2008), which are applied in this model (using "method C"). Garrod et al. (2009) showed that this method produces a very close approximation to the results of exact Monte Carlo methods, using a large chemical network under a range of physical conditions. \\

\subsection{Chemistry}

The chemical network used in this model is based on that of GWH and the subsequent additions of Belloche et al. (2009), which featured a treatment for nitriles, including amino acetonitrile (NH$_2$CH$_2$CN). These networks assumed grain-surface radical--radical formation mechanisms for newly-added complex organic molecules, as well as destruction mechanisms such as grain-surface and gas-phase photodissociation (by both the ambient and cosmic ray-induced UV fields), gas-phase ion-molecule reactions (including protonation) and the related electronic recombinations of the products, as well as accretion from the gas-phase onto the grain surfaces, and thermal and non-thermal desorption from the grains. The initial elemental abundances used are shown in Table \ref{tab-init}. \cite{Laas11} constructed a network of reactions to treat the formation of {\em trans}-methyl formate, the higher energy conformer of the commonly observed {\em cis}-methyl formate; however, that study found no significant effect on the abundances of species other than {\em trans}-MF, so the omission of the related chemistry is not expected to have any effect on the models presented here. 

Following the same methodology, formation and destruction processes are included for glycine and a suite of related molecules, including glycinal, propanal and propionic acid. Radical--radical reactions for a selection of these species are shown in Table \ref{tab-surf}. Whereas in the previous models, a single complex molecule was typically assumed to be the only possible product of the addition of key grain-surface radicals, here alternative product branches are included, for the case where such a reaction could also result in a pair of stable products, and where the reaction is also exothermic. Thus, for the reaction HCO + CH$_3$O $\rightarrow$ HCOOCH$_3$, which was found by GWH to be the predominant formation route for methyl formate, alternative product branches of [CO + CH$_3$OH] and [H$_2$CO + H$_2$CO] are now also included. The branching ratios are in general poorly defined, and are likely to be significantly different between the gas- and solid-phase processes; statistical ratios are therefore adopted (e.g. 1:1:1 for HCO + CH$_3$O). In addition, the same such two-product reactions are now also included in the gas-phase chemistry, using measured rates\footnotemark \ where available; otherwise, a uniform second-order rate coefficient of $10^{-11}$ cm$^3$ s$^{-1}$ is assumed. Such processes may become important in the late, high-temperature stage of hot-core evolution, during which the evaporation, protonation and subsequent electronic recombination may lead to significant quantities of molecular radicals being present in the gas phase. However, the single-product branches are omitted from the gas-phase chemistry, because of their negligible rates, due to the absence of a grain surface to carry away excess energy and thereby stabilize the product. 

\footnotetext{NIST Chemical Kinetics Database, www.kinetics.nist.gov/kinetics}

Following previous models, grain-surface hydrogen-abstraction reactions are added for reactions between the new complex species and the most significant grain-surface radicals such as H, OH, CH$_3$, CH$_2$OH, CH$_3$O and NH$_2$. Table \ref{tab-act} shows activation-energy barrier information used in the model, for a selection of (primarily hydrogen-abstraction) reactions occurring on the grains. For atomic-hydrogen reactions relating to methanol formation, for which more complex fits to theoretical rates are employed, the simpler rectangular-barrier parameter fits are quoted, as for all other barrier-mediated reactions. The method of Hasegawa et al. (1992) is used to calculate the tunneling and thermal rates associated with these parameters, adopting the faster of the two for any given temperature. For the majority of reactions, only gas-phase data are available, in which case the activation energy derived from fits to the lowest-temperature data-sets available are adopted for the surface/mantle reactions, assuming a barrier width of 1 \AA. In cases where no data exist, the activation energy, $E_{\mathrm{A}}$, is estimated using the Evans-Polanyi relationship, which states that the difference in $E_{\mathrm{A}}$ between reactions of a similar type is proportional to the difference between the enthalpy changes associated with each reaction. Parameters for the Evans-Polanyi relationship are obtained from Dean \& Bozzelli (1999), or from fits using similar reactions, indicated in Table \ref{tab-act}. Where the enthalpy change of reaction is unknown, the activation energy of a similar reaction is used directly. There is evidence that the presence of a surrounding water-ice matrix may alter activation energy barriers for specific product branches (e.g. Goumans 2011, in the case of formaldehyde-related chemistry), or lower the internal barriers to the structural re-arrangement of a product (e.g. Duvernay et al. 2005, in the case of the isomerization of cyanamide). It is therefore plausible that some of the gas-phase activation energies used in this model over-estimate the barriers in the solid phase; however, such effects are difficult to quantify without a specific computational or experimental study of each solid-phase reaction and its low-temperature tunneling characteristics.

For self-consistency, all reactions of surface radicals with complex species are now assumed to result in hydrogen abstraction, in contrast to GWH, who included a small selection of alternative substitution reactions for aldehyde group-bearing species. These reactions typically provided only minor destruction or formation routes for the species involved, and their removal has no significant influence over the glycine chemistry described below.

All processes related to grain-surface chemistry outlined above are likewise incorporated into the chemistry of the bulk ice, with identical activation energies (where present), while their reaction kinetics are instead defined by the bulk-ice diffusion parameters discussed in Section 2.1. Photodissociation of ice mantles is also included to all depths, without attenuation, in line with previous models. This is a reasonable first-order approximation; using the photon absorption cross-section of amorphous water ice for the first absorption band measured by Mason et al. (2006), Andersson \& van Dishoeck (2008) calculate an absorption probability per photon of 0.007 per monolayer of water ice. The maximum quantity of water ice formed on the grains in this model corresponds to $\sim$114 monolayers; this would result in a UV field strength at the deepest ice layer of $\sim$45\% of the value at the surface. This corresponds to a mean value of 69\% of the assumed field strength, for the ice mantle as a whole. However, the ices formed on interstellar dust-grain surfaces constitute a mixture of various molecular species, and the absorption cross-sections for many of these are currently unknown or poorly defined.

No desorption processes are allowed from the bulk ice; thus desorption of these species must proceed via a prior transfer of material from mantle to ice surface. Reaction--diffusion competition is also included for reactions mediated by activation-energy barriers (following G\&P), for both surface and bulk-ice reactions; this is perhaps the largest departure from the methods of GWH, as this change can produce reaction efficiencies many orders of magnitude different from the simple Arrhenius treatment of previous models.

Surface desorption processes included are thermal evaporation (as defined by the binding/desorption energy, $E_{\mathrm{des}}$), reactive desorption as described by Garrod et al. (2007), and photo-desorption using measured rates determined by Oberg et al. (2009c,d); following G\&P, all surface species are assumed to photodesorb with an efficiency of $10^{-3}$, in the case where values are not provided by those authors.

The indirect mechanism for CO$_2$ formation proposed by G\&P is also included for reactions both on the surfaces and within the ice mantles.

The reproduction of the high observed abundance ($\sim$$10^{-8}$ with respect to H$_2$) of methyl formate (HCOOCH$_3$) in hot cores has proved somewhat problematic in previous astrochemical models. Garrod \& Herbst (2006) showed that grain-surface formation mechanisms could produce adequate quantities of methyl formate (MF) to explain observations; however, in some cases it was found (Garrod et al. 2008; Belloche et al. 2009) that the evaporation of MF at relatively low temperatures could result (dependent on the warm-up timescale) in only a small fraction of this material surviving the transition to the gas-phase. Belloche et al. (2009) adopted a higher binding energy for MF (5200 K vs. 4100 K), citing the importance of hydrogen bonding in the estimation of this value; they found that this value produced results appropriate to observational abundance values for methyl formate. This value is also adopted here. Table \ref{tab-bind} shows a selection of other binding energies used in the present model.

\subsection{Methanol chemistry}

As found by GWH and Laas et al. (2011), the rates and branching ratios of methanol photodissociation in the ice mantles are crucial to the production rates and ratios of various complex organic molecules detected in hot cores, as the ice-mantle methanol provides a large supply of structurally-complex organic material. 

The present model uses the methanol branching ratios derived from the experiments of Oberg et al. (2009a) for products [CH$_2$OH+H]:[CH$_3$O+H]:[CH$_3$+OH] of 5:1:1. In the absence of other information, these values are applied to both the direct UV photodissociation by the interstellar radiation field, and to photodissociation by the secondary UV photons produced by cosmic rays, the latter of which is most important at the high visual extinctions of the hot-core phase. These ratios are applied throughout to the ice and gas phases, although, as found by Laas et al. (2011), the importance of gas-phase methanol photodissociation to complex molecule production is minimal.

Activation-energy treatments for CO and H$_2$CO use best-fit expressions to uni-molecular and bi-molecular rates recently acquired through quantum transition state theory (Goumans, priv. comm.; Andersson, priv. comm.). Hydrogen abstraction from H$_2$CO and CH$_3$OH are included, mediated by activation energy barriers, while hydrogen abstraction from HCO and CH$_3$O/CH$_2$OH are also allowed (producing H$_2$ + CO/H$_2$CO), assuming branching ratios of 1:1 with the associated hydrogen-addition channels (i.e. no activation energy).

\subsection{Glycine chemistry}

\subsubsection{Grain-surface/mantle chemistry}

A number of potential routes have been suggested in the literature for the formation of glycine in interstellar ices, based on laboratory evidence or theoretical considerations. Experiments by Bernstein et al. (2002) and Mu{\~n}oz Caro et al. (2002) showed that UV irradiation of interstellar ice analogs containing H$_2$O, CH$_3$OH, CO, CO$_2$, NH$_3$ and/or HCN could produce significant quantities of amino acids, including glycine. Sorrell (2001) had suggested a route involving the reaction of the NH$_2$ radical with acetic acid (CH$_3$COOH), itself formed through the addition of CH$_3$ and HOCO, while Woon (2002) suggested that the addition reaction between radicals NH$_2$CH$_2$ and HOCO could explain the formation of glycine, with these species produced in the ices by successive hydrogenation of HCN and by the addition of CO to OH, respectively. Experimental work by Holtom et al. (2005) involving the electron bombardment of CO$_2$/NH$_2$CH$_3$ ices confirmed the viability of Woon's mechanism. Elsila et al. (2007) found Woon's mechanism to be consistent with the origins of the nitrogen and central carbon in the glycine molecule, while the origin of the acidic carbon was more consistent with the formation of glycine occurring via a nitrile precursor, i.e. amino acetonitrile (NH$_2$CH$_2$CN).

Here, we include a number of new radical-addition reactions relating to glycine and similar species. Each hydrogenation step suggested by Woon (2002) is included, although all but the final addition of NH$_2$CH$_2$ + HOCO was already present in the network. However, other routes are also included, such as the addition reaction NH$_2$CH$_2$ + HCO $\rightarrow$ NH$_2$CH$_2$CHO (glycinal), whose product may (for example) be photodissociated to produce the radical NH$_2$CH$_2$CO, which may further react with OH to produce glycine. There are four final processes that lead to glycine in this network,
\begin{eqnarray}
&\mathrm{H} + \mathrm{NH}\mathrm{CH}_{2}\mathrm{COOH} \rightarrow \mathrm{NH}_{2}\mathrm{CH}_{2}\mathrm{COOH} \\
&\mathrm{NH}_{2} + \mathrm{CH}_{2}\mathrm{COOH} \rightarrow \mathrm{NH}_{2}\mathrm{CH}_{2}\mathrm{COOH} \\
&\mathrm{NH}_{2}\mathrm{CH}_{2} + \mathrm{HOCO} \rightarrow \mathrm{NH}_{2}\mathrm{CH}_{2}\mathrm{COOH} \\
&\mathrm{NH}_{2}\mathrm{CH}_{2}\mathrm{CO} + \mathrm{OH} \rightarrow \mathrm{NH}_{2}\mathrm{CH}_{2}\mathrm{COOH}
\end{eqnarray}
although there are several ways in which each radical may come to be formed, either through chemical reaction or by photodissociation of a stable molecule. Reaction (5) is also complemented by exothermic radical-radical reactions with species (e.g. HCO) that may donate a hydrogen atom to form two stable molecules, as described in section 2.2. Reaction (5) typically follows the photodissociation of, or H-abstraction from, glycine itself; reactions (6)-(8) are thus the major potential formation routes.

Woon suggests the formation of HOCO by the reaction of CO and OH. This process is difficult to model accurately, due to the complex nature of the transition states; the reaction may also lead to the formation of CO$_2$ + H (see G\&P, and references therein), or may dissociate back to CO + OH, dependent on the rapidity of energy loss to the surface. As such, the OH + CO reaction is modeled separately for each set of products, assuming an activation energy barrier of 2500 K for the HOCO branch, following the approach of GWH. In practice, the formation of HOCO in the model is dominated by the photodissociation of or chemical hydrogen abstraction from formic acid (HCOOH) in the ice, which is formed predominantly by HCO + OH additions (see GWH).

Also included in the model is a selection of species similar to glycine, but with alternative functional groups. For example, glycinal (NH$_2$CH$_2$CHO) and propionic acid (C$_2$H$_5$COOH) are modeled, through analogous reactions to those of glycine, while new dust-grain formation routes are added for propionaldehyde (or propanal, C$_2$H$_5$CHO), which has been detected toward Sgr B2(N) (Hollis et al. 2004). The inclusion of these other species ensures that the importance of reactions leading to glycine formation is not over-estimated due to the omission of competing reaction pathways.

\subsubsection{Gas-phase chemistry}

As for other complex species, the chemical network includes gas-phase destruction routes for glycine that consist of UV photodissociation, either by the cosmic ray-induced or standard interstellar radiation field, or by reactions with atomic or simple molecular ions. Table \ref{tab-gas} shows the newly-included reactions for glycine and its related gas-phase products. Electronic recombination of protonated glycine results in the re-formation of glycine with an efficiency of 5 \%, but all the glycine in the gas-phase is originally formed on the grains. Consideration of a potential gas-phase formation route is given in the models of Section 5.

\subsection{Physical model}

The physical model used here broadly follows that of G\&H and later papers, in which a collapse phase is followed by a static warm-up phase. In the first phase, the density increases from $n_{\mathrm{H}}=3 \times 10^{3}$ to $10^7$ cm$^{-3}$, under the free-fall collapse expression given by Nejad et al. (1990). However, in contrast to previous hot-core models, while the gas temperature is held at a constant 10 K, the temperature of the dust varies with visual extinction, according to equation (17) of G\&P, which describes the heating of the dust by the interstellar radiation field, assuming a dust-grain radius of 0.1 $\mu$m. At the initial visual extinction of $A_{V}=2$, the dust temperature is $\sim$16 K; during collapse, this falls to a minimum temperature of 8 K. 

During the subsequent warm-up phase, the core is heated from 8 -- 400 K, with density held at $10^7$ cm$^{-3}$. Three warm-up timescales are adopted, which are the same as previous models, in so far as they reach 200 K at the same point in time (Table \ref{tab-mod}). However, the warm-up is extended beyond this time, until 400 K is reached, so that the later evaporation and gas-phase behavior of more strongly-bound species, such as glycine, may be studied over a longer period. During the warm-up phase, gas and dust temperatures are assumed to be well coupled due to the high density; the gas kinetic temperature follows that of the dust grains when $T_{\mathrm{dust}}>$10 K. Table \ref{tab-mod} shows the details of the three warm-up phase physical models.

Table \ref{tab-ice} shows the composition of the dust-grain ices at the end of the collapse phase (corresponding to the starting values for each of the following warm-up models), as well as observational values for comparison.

\section{Results}

Figures 1, 2 and 3 show the behavior of fractional abundances with respect to total hydrogen ($n_{\mathrm{H}}$) of a selection of simple and complex molecules, for the {\em fast}, {\em medium} and {\em slow} warm-up models, respectively. Peak gas-phase abundances are shown in Table \ref{tab-output}, along with the corresponding system temperatures for each peak value, which frequently correspond to the temperature of peak evaporation for species formed on the dust grains. 

In Figs. 1 -- 3, panel (a) shows some of the main ice constituents and a representative ion, HCO$^+$. Panels (b) and (c) show some complex organics typically observed toward hot cores. Panel (d) shows some commonly observed nitrogen-bearing species as well as the as-yet undetected hydroxylamine (NH$_2$OH). Panel (e) shows a selection of complex species containing a carbonyl group (C=O), including acetone, (CH$_3$)$_2$CO. Panel (f) shows complex species derived from the larger photo-fragments of methanol. Panel (g) shows a selection of complex organics bearing an amine group (NH$_2$), while panel (h) shows glycine itself (NH$_2$CH$_2$COOH) along with several similar or related molecules. Not all species shown in Table \ref{tab-output} are plotted in Figs. 1 -- 3.

The results for a selection of important species and their progenitors are discussed in detail below. Due to the large number of molecules included in the model, the results for many molecules are presented in Figs. 1 -- 3 and Table \ref{tab-output} without comment. For a description of the most important reaction mechanisms relating to these species, the reader is referred to the analysis presented by Garrod et al. (2008); however, some general differences between the current and previous models are discussed in section 6.

\subsection{Glycine and related species}

\subsubsection{Glycinal}

Glycinal (NH$_2$CH$_2$CHO) is the amino aldehyde corresponding to the amino acid glycine. Its formation in these chemical models, shown in panel (h) of Figs. 1 -- 3, begins at around 25 K, when the radical HCO becomes mobile and reacts with CH$_2$NH$_2$. The latter species is formed by the hydrogenation of CH$_2$NH, which builds up at around 15 K via the addition of NH and CH$_2$ radicals. The HCO radical is mainly formed by the abstraction of hydrogen from formaldehyde (H$_2$CO) by OH, which is a barrierless process; this occurs until formaldehyde evaporates at around 40 K (see G\&H, GWH, and section 3.2, below). A later and somewhat stronger glycinal formation route operates at $\sim$40 K, when NH$_2$ radicals add to CH$_2$CHO radicals formed by the photodissociation of acetaldehyde (CH$_3$CHO).

Glycinal formation thus occurs at rather lower temperatures than that of glycine (see below). However, both are at least partially dependent on the earlier formation of acetaldehyde at lower temperatures to provide the CH$_2$CO structure present in both. Peak gas-phase glycinal abundances are a few $10^{-9}$ to $10^{-8}$ with respect to total hydrogen; it is thus of similar abundance to the more prevalent observed organic molecules in hot cores. No mm/sub-mm spectral information is apparently available for glycinal at present.

\subsubsection{Glycine}

Glycine formed on the grains is seen to be released into the gas phase at a temperature of around 200 K, attaining a peak gas-phase abundance of $8.6 \times 10^{-11}$, $4.6 \times 10^{-10}$, and $8.1 \times 10^{-9}$ with respect to total hydrogen (see Table \ref{tab-output} and Figs. 1--3 panel h), for the {\em fast}, {\em medium}, and {\em slow} warm-up timescales, respectively.

The three main production routes included for glycine (Eqs. 6 -- 8) are all active, and all contribute a proportion of the glycine later released into the gas phase. Significant dust-grain surface/mantle formation of glycine begins at temperatures of 40 -- 50 K, and may continue up to around 120 K.

For the {\em medium} warm-up timescale case, at around 40 K the OH radicals in the ice, produced by the photodissociation of water molecules, are sufficiently mobile to find glycinal (NH$_2$CH$_2$CHO), which is abundant at around $10^{-9} - 10^{-8}$ with respect to total hydrogen. The OH radical easily abstracts a hydrogen atom from glycinal; the activation energy barrier is sufficiently low (591 K) compared to the diffusion barrier for OH that the reaction occurs before OH has an opportunity to diffuse away from its reaction partner. (This effect indeed confers an efficiency close to unity on most such OH-initiated hydrogen-abstraction reactions). The resulting radical, NH$_2$CH$_2$CO, then reacts with another OH to produce glycine (Eq. 8).

At temperatures of 55 K and above, much greater quantities of glycine are formed; hydrogen abstraction from acetic acid (CH$_3$COOH) by NH$_2$ radicals within the ice mantles and the subsequent addition of the resultant CH$_2$COOH radicals to NH$_2$ produces glycine (Eq. 6). The NH$_2$ radicals are primarily formed by hydrogen abstraction from ammonia (NH$_3$) by OH radicals, which are themselves formed by the photodissociation of water ice. The increased importance to glycine formation of the acetic acid-related path at these temperatures is caused by the greater abundance of NH$_2$ radicals that results from the evaporation of HNCO molecules from the ices, which up until that time represent a major reaction partner for NH$_2$. The release of HNCO into the gas phase may be seen in Figs. 1 -- 3, panel (d). However, over the 55 -- 75 K range, Eq. (8) still dominates Eq. (6) by a factor of 2 -- 3, as the greater availability of NH$_2$ allows the NH$_2$CH$_2$CO radical to be formed by hydrogen abstraction from glycinal by NH$_2$ as well as OH.

At around 75 K, an ice {\em surface} (as opposed to {\em mantle}) process becomes the strongest contributor to glycine formation; hydrogen abstraction from CH$_3$NH$_2$ (methylamine) by OH radicals produces CH$_2$NH$_2$, while abstraction from HCOOH (formic acid) by OH produces HOCO. These react to produce glycine (Eq. 7) on the ice surface, by the route suggested by Woon (2002). This mechanism then competes roughly equally with the reaction of CH$_2$COOH and NH$_2$ on the ice surface (Eq. 6) to be the main producer of glycine, up to $\sim$90 K. By this point, most glycine formation is taking place on the surfaces of the ices, rather than within the ice mantles. This is caused by the increasing rate of mantle-to-surface diffusion among less strongly-bound species, including NH$_2$, as the temperature rises, coupled with their commensurately faster reaction rates on the ice surface. The more strongly-bound products of these reactions, such as glycine itself, are ``swapped'' back into the deeper mantle following their formation, as they are exchanged with more mobile, volatile species.

Between around 90 -- 120 K, the reaction of NH$_2$ and CH$_2$COOH (Eq. 6) again takes over as the dominant route to glycine on the grains, as CH$_3$NH$_2$ is gradually lost to the gas phase, rendering Eq. (7) far less effective. 

The majority of the glycine ultimately released into the gas phase is formed during the 90 -- 100 K period. During this time, Eq. (6) dominates Eq. (7) by a factor of $\sim$2.

At around 120 K (in this and all models), ammonia (NH$_3$), formic acid (HCOOH), and acetic acid (CH$_3$COOH) evaporate fully from the grains, closing down the main glycine formation mechanisms. From this point on, glycine begins to migrate from the bulk ice mantle into the surface, while suffering a moderate degree of attrition by hydrogen abstraction and photodissociation -- the products of which typically evaporate.

For the {\em fast} warm-up model, the temperature dependence of the formation mechanisms is broadly similar; however, the importance of each route is different, due to the differing quantities of precursor molecules available. Most importantly, the abundance of acetic acid (CH$_3$COOH), shown in panel (e), is around one order of magnitude lower than in the {\em medium} timescale model. As found by Garrod et al. (2008), in these models acetic acid is formed predominantly by the addition of OH to CH$_3$CO, the product of the abstraction of hydrogen from acetaldehyde (CH$_3$CHO) by OH itself. The {\em fast} warm-up model allows a shorter time period for this process, before acetaldehyde evaporates from the grains, resulting in the lower acetic acid abundance. This in turn renders the production of glycine via Eq. (6) generally a factor of a few slower than Eqs. (7) \& (8), which, overall, provide approximately equal amounts of glycine production in this model.

For the {\em slow} warm-up model, the behavior of glycine production is much more similar to that of the {\em medium} case; however, here the NH$_2$ + CH$_2$COOH route is strongly dominant over the 60 -- 120 K range during which glycine production is important. This is caused by the greater time period for the production of acetic acid. Production via Eq. (8) is also less significant than in the two shorter-timescale models, due to the lower abundance of glycinal in the ices.

In all models, the gas-phase destruction of glycine is entirely dominated by protonation and dissociative electronic recombination, as is the case for the majority of complex organics (see also GWH); photodissociation has only a small effect, as the only UV radiation field capable of influencing the chemistry during the warm-up phase is that induced by cosmic rays.

These models produce a low to moderate abundance of glycine in the gas phase, in comparison with other complex species produced in the models. The peak gas-phase abundance of glycine is also very similar to that of acetic acid in each model, albeit with a different temperature of evaporation. The question of whether this glycine may be detectable is addressed in Section 5.

\subsubsection{Amino acetonitrile}

Amino acetonitrile (NH$_2$CH$_2$CN), or AAN, shown in panel (h), has been suggested as a precursor to glycine (Elsila et al. 2007). It is formed in these models with a peak gas-phase fractional abundance of a few $10^{-9}$ to $10^{-8}$, consistently for each timescale model. This is around a factor of two higher than the values produced in the model of Belloche et al. (2009), and around a factor of 10 greater than the value derived from observations toward Sgr B2(N) by Belloche et al. (2008).

In these models, AAN is formed primarily by the addition of NH$_2$ to the radical CH$_2$CN, which itself derives from hydrogen abstraction from CH$_3$CN molecules by NH$_2$ at around 60 -- 80 K, producing an abundance of around $10^{-9}$ on the grains. Later, at around 90 K, the evaporation of CH$_3$CN from the grains results in its gas-phase protonation and recombination with electrons, producing CH$_2$CN with an efficiency of 50 \%. This radical re-accretes onto the grains, to react with NH$_2$, which produces an AAN fractional abundance of around $10^{-8}$ on the grains. However, in light of recent measurements of methanol recombination branching ratios (e.g. Geppert et al. 2006), it is unclear whether such a high efficiency of CH$_2$CN production by gas-phase electronic recombination is realistic. A value of 5 \%, as adopted in these models for the recombination of more complex species, would render the earlier formation of AAN by purely grain-surface chemistry the dominant mechanism, and would produce abundances close to the values observed toward Sgr B2(N).

\subsubsection{Propionaldehyde and propionic acid}

Propionaldehyde (or propanal, C$_2$H$_5$CHO) and propionic acid (or propanoic acid, C$_2$H$_5$COOH) are structurally similar to glycinal and glycine, and neither have been treated in chemical models previously. Propionaldehyde is formed in great abundance at low temperatures, by the addition of HCO and C$_2$H$_5$ radicals on the dust-grain ice surfaces; abundances as high as $10^{-7}$ with respect to total hydrogen are formed in the {\em slow} warm-up model. Production is most rapid at around 30 K, when the evaporation of grain-surface methane (CH$_4$, panel a) is fastest. This results in a significant increase in the abundance of gas-phase acetylene (C$_2$H$_2$), which accretes onto the dust grains, where it is rapidly hydrogenated to C$_2$H$_5$, thence reacting to form propionaldehyde. Formation on the ice surface at this time is sufficiently rapid that so-called reactive desorption, whereby the energy released from the exothermic formation of a surface molecule results in desorption (e.g. Garrod et al. 2007), produces a gas-phase peak in the abundance of propionaldehyde of approximately $10^{-12}$ $n_{\mathrm{H}}$. This period of carbon-rich gas-phase chemistry associated with methane evaporation is often labeled warm carbon-chain chemistry (WCCC), and has been identified observationally by Sakai et al. (2008); however, it is unlikely that the enhancement in propionaldehyde seen in the models would be observable at the low abundances predicted by these models. The large quantity that remains on the grain surfaces later evaporates at around 120 -- 130 K, producing gas-phase peak fractional abundances of around $10^{-8}$. This molecule was detected by Hollis et al. (2004) toward Sgr B2(N); however, no specific abundance was determined.

Propionic acid (C$_2$H$_5$COOH) is formed on the grains much later than propionaldehyde; most production occurs from around 70 K onwards, by the addition of OH and C$_2$H$_5$CO radicals on the ice surface. The latter species is formed primarily through hydrogen abstraction from propionaldehyde by NH$_2$ in the ice. Thus, propionic acid production appears to be directly dependent on the prior formation of its associated aldehyde.

\subsection{Other complex molecules}

\subsubsection{Methanol and formaldehyde}

Methanol (CH$_3$OH) and formaldehyde (H$_2$CO), shown in Figs. 1--3 panel (b), are both present in the dust-grain ice mantles in large quantities, following their formation at low temperature during the collapse phase (see Table \ref{tab-ice}). Formaldehyde remains the primary source of HCO radicals in the ices, due to hydrogen abstraction by OH, until it evaporates at around 40 K, producing a significant gas-phase peak -- see G\&H and GWH. This feeds the gas-phase formation of other species, most notably formic acid and methanol.

Methanol abundance in the ice is essentially static until it eventually evaporates at around 120 K. But methanol nevertheless provides the majority of organic structure for the formation of more complex molecules, in the form of methyl (CH$_3$), methoxy (CH$_3$O) and hydroxymethyl groups (CH$_2$OH). These are formed primarily by photodissociation caused by the cosmic ray-induced UV field that penetrates the hot core. Above $\sim$45 K, OH radicals become mobile, and hydrogen abstraction to form CH$_2$OH becomes the dominant destruction mechanism for methanol, producing around twice as much CH$_2$OH as the photodissociation route. 

The abundance of methanol in the ices is little affected by the warm-up timescales of each model. The fall in the gas-phase abundances is more significant, but the apparent differences between each timescale model are mainly due to the longer timescales over which the chemistry of the slower warm-up models is simulated, rather than to a difference in the rate of destruction. The temperature--abundance relationships may, however, be important to the interpretation of observationally-determined abundances and rotational temperatures. The apparently precipitous fall in the gas-phase abundances of various other molecules in the {\em slow} warm-up model is also largely attributable to the longer timescales adopted in that model. 

The eventual evaporation of methanol also indirectly produces a high-temperature peak in formaldehyde abundance; the reaction of methanol with protonated methanol produces protonated dimethyl ether (CH$_3$OCH$_4$$^+$), one of whose electronic recombination products is assumed to be formaldehyde in this model. The branching products of this process are not well constrained, {\em pace} Hamberg et al. (2010), who indeed do not identify H$_2$CO as a likely product. The secondary peak in formaldehyde abundance may at least be considered unreliable, due to these uncertainties.

\subsubsection{Methyl formate and glycolaldehyde}

The peak abundance of methyl formate (MF, HCOOCH$_3$), shown in Figs. 1--3 panel (b), is in line with that expected from observations (around $10^{-8} n_{\mathrm{H}}$), except for the {\em slow} warm-up model. The evaporation of methyl formate prior to that of methanol and various other species leaves it vulnerable to destructive reactions with molecular ions, whose abundance would otherwise be lessened by the presence of other reaction partners. The long timescale between methyl formate and methanol evaporation in the {\em slow} model therefore leads to significant methyl formate destruction; a similar result was noted by GWH and Belloche et al. (2009), in the case of a lower assumed MF binding energy.

Methyl formate is formed entirely on the dust grains in this model, by the reaction HCO + CH$_3$O. Production is mainly within the ice mantles, peaking at 25 K, although significant production also occurs on the ice surface.

Glycolaldehyde (CH$_2$(OH)CHO), shown in Figs. 1--3 panel (c), is formed similarly to its a structural isomer methyl formate, via HCO + CH$_2$OH. Its gas-phase and ice abundance is greater than that of MF, due to the more rapid production of CH$_2$OH over CH$_3$O -- by a factor of 5 (for photodissociation) in this model. Observations of protostellar envelopes, however, do not demonstrate this large abundance of glycolaldehyde; the recent detection by J{\o}rgensen et al. (2012) toward the Class 0 protostellar binary IRAS 16293-2422, using the ALMA telescope, indicated a HCOOCH$_3$:CH$_2$(OH)CHO ratio of 10 -- 15 (although the observational glycolaldehyde rotational temperature of 200 -- 300 K {\em is} consistent with the present model results).

The discrepancy may be related to the assumed branching ratios of the reactions between HCO and either CH$_3$O or CH$_2$OH:
\begin{mathletters}
\begin{eqnarray}
\mathrm{HCO} + \mathrm{CH}_{3}\mathrm{O} & \rightarrow \mathrm{HCOOCH}_{3}                                    &  (-423 \ \mathrm{kJ/mol}) \\
                                                                      & \rightarrow \mathrm{CO} + \mathrm{CH}_{3}\mathrm{OH}  & (-372 \ \mathrm{kJ/mol}) \\
                                                                      & \rightarrow \mathrm{H}_{2}\mathrm{CO} + \mathrm{H}_{2}\mathrm{CO}  & (-292  \ \mathrm{kJ/mol}) 
\end{eqnarray}
\end{mathletters}
and
\begin{mathletters}
\begin{eqnarray}
\mathrm{HCO} + \mathrm{CH}_{2}\mathrm{OH} & \rightarrow \mathrm{CH}_{2}\mathrm{(OH)CHO}  &  (-290 \ \mathrm{kJ/mol}) \\
                                                                      & \rightarrow \mathrm{CO} + \mathrm{CH}_{3}\mathrm{OH}  & (-276 \ \mathrm{kJ/mol}) \\
                                                                      & \rightarrow \mathrm{H}_{2}\mathrm{CO} + \mathrm{H}_{2}\mathrm{CO}  & (-197  \ \mathrm{kJ/mol})
\end{eqnarray}
\end{mathletters}
\noindent (The enthalpy change associated with each reaction, $\Delta H_{f}^{0}$, is indicated in brackets). The branching of each set of reactions is assumed to be [1:1:1] in these models. Each branch of equations (9) and (10) is exothermic, and branches (a) and (b) would require minimal internal re-arrangement. The single-product routes in each case are the most exothermic; however, in the case of glycolaldehyde formation, the enthalpy change related to reaction branch (10b) is close to that of (10a). While the assumed efficiency of branch (c) (i.e parity with the other two branches) may be over-estimated in both equations (9) and (10), the production of CO and CH$_3$OH (10b) may be very competitive with the formation of glycolaldehyde (10a), while it may be less so in the case of methyl formate.

It thus appears plausible, by this simplistic analysis, that a difference in the efficiencies of formation of MF and glycolaldehyde via their primary grain-surface/ice-mantle formation routes could account for the order-of-magnitude discrepancy in observational abundances. However, a detailed theoretical determination of the branching behavior of these reactions in the solid phase would be necessary to confirm this theory. Alternative branching ratios in the photodissociation of methanol, to produce a significant excess of CH$_3$O over CH$_2$OH is also possible, although the current -- and only -- estimate of these values for solid-state methanol, based on experimental rate-fitting, suggests precisely the opposite relationship (CH$_3$O:CH$_2$OH of 1:5; {\"O}berg et al. 2009a).

A decrease in the efficiency of glycolaldehyde production as described above would be unlikely to affect significantly the abundances of other {\em observed} species, although the production of postulated molecules related to glycolaldehyde, such as dihydroxyacetone (HOCH$_2$COCH$_2$OH) could be affected. The destruction rates of the HCO and CH$_2$OH radicals would remain the same, while the increased re-formation of CO and CH$_3$OH resulting from the alternative product branch would have only a minor effect on the already large abundances of those species. 

The glycine formation routes included in this model are sufficiently independent from the glycolaldehyde chemistry that glycine production would also be unaffected by the adoption of an alternative efficiency for glycolaldehyde formation. Similarly, while acetic acid (CH$_3$COOH) is a structural isomer of both glycolaldehyde and methyl formate, its different structure means that its formation routes are not directly related to those of its isomers; the choice of treatment for glycolaldehyde would there have little effect on its own abundance or, through it, that of glycine. The chemical behavior of acetic acid is discussed above, in section 3.1.1.

\subsubsection{Formic acid}

Formic acid (HCOOH), shown in Figs. 1--3 panel (b), is formed to a large degree in the gas phase, especially for the {\em fast} and {\em medium} warm-up models; this occurs when formaldehyde (H$_2$CO) evaporates at $\sim$40 K, reacting with OH to form HCOOH and atomic hydrogen. The gas-phase formic acid accretes back onto the grains, and does not evaporate until a temperature of around 120 K is reached. A similar or somewhat smaller quantity of the formic acid that ultimately evaporates from the grains is formed at low temperature within the ice mantles by the addition of HCO and OH, as HCO becomes mobile.

As such, it is unclear whether the formic acid observed toward hot cores originates in the gas phase or on the grains. As discussed above for the case of methyl formate and glycolaldehyde production, a plausible alternative product branch for the reaction of HCO and OH exists, yielding CO and H$_2$O (rather than HCOOH); indeed, this branch is somewhat more exothermic than the single-product branch. Both branches were included in the model, at a 1:1 efficiency ratio. The true efficiency for HCOOH formation could therefore be twice as large, or significantly smaller than assumed. The rate of the gas-phase reaction (OH + H$_2$CO $\rightarrow$ HCOOH +H) is reasonably well known, albeit only around room temperature (Yetter et al. 1989, and references therein); but the amount of formic acid formed as a result is strongly dependent on the amount of formaldehyde that is released into the gas-phase from the grains.

\subsubsection{Formamide}

The behavior of the amino aldehyde formamide (NH$_2$CHO), shown in Figs. 1--3 panel (d), is somewhat different to that presented in the models of GWH. In this case, the quantity formed in the ices, via the addition of HCO and NH$_2$ radicals, is somewhat larger than previously found. However, its peak gas-phase abundance is lower, and falls over time, rather than rising as was found by GWH. The latter change in behavior is due to the removal of a single gas-phase reaction from the network: NH$_2$ + H$_2$CO $\rightarrow$ NH$_2$CHO + H. This reaction was present in previous networks with a generic collisional rate coefficient of $10^{-10}$ cm$^{3}$ s$^{-1}$, but it is likely to have an activation energy barrier (based on the barrier of 5.89 kcal/mol ($\sim$3000 K) calculated by Li \& Lu (2002) for the more exothermic product branch NH$_3$ + HCO). The analogous reaction OH + H$_2$CO $\rightarrow$ HCOOH +H, discussed above, has a measured rate of $2 \times 10^{-12}$ cm$^{3}$ s$^{-1}$, corresponding to a maximum 2\% efficiency in the branching ratio compared to the alternative H$_2$O + HCO branch. A theoretical study (Li \& Lu, 2002) of the NH$_2$ + H$_2$CO $\rightarrow$ NH$_3$ + HCO reaction suggested a rate coefficient of $5.25 \times 10^{-17}$ cm$^{3}$ s$^{-1}$. For these reasons, the NH$_2$ + H$_2$CO reaction was removed entirely.

\section{Gas-phase routes to glycine formation}

While laboratory experiments have shown success in producing glycine and other amino acids by the photolysis of organic ice mixtures (Bernstein et al. 2002; Mu{\~n}oz Caro et al. 2002; Holtom et al. 2005; Elsila et al. 2007), gas-phase mechanisms for amino-acid production have also been considered. Blagojevic et al. (2003) demonstrated that reaction between acetic acid (CH$_3$COOH) and ionized or protonated hydroxylamine (NH$_{2,3}$OH$^+$) can result in the formation of ionized or protonated glycine (NH$_{2,3}$CH$_2$COOH$^+$) and water. Similar schemes involving propionic (propanoic) acid were found to produce the equivalent forms of alanine. The authors estimated rate coefficients of $>10^{-12}$ cm$^3$ s$^{-1}$ for each of these processes.

The most probable source of NH$_{2,3}$OH$^+$ ions in hot cores would be NH$_2$OH, either by photoionization, electron transfer with atomic ions such as He$^+$ and C$^+$, or protonation by proton donors such as H$_3$$^+$, H$_3$O$^+$ or HCO$^+$, all of which processes are already present in the chemical network. Protonation by CH$_5$$^+$, as suggested by Snow et al. (2007), is not included; however, H$_3$$^+$ and H$_3$O$^+$ are around 2 orders of magnitude more abundant than CH$_5$$^+$ in the models, while the protonation rate coefficients in each case are typically very similar. Protonation rates of hydroxylamine by HCO$^+$, H$_3$$^+$ and H$_3$O$^+$ of $(0.59$, $1.39$ and $6.68) \times 10^{-9} (T/300 \ \mathrm{K})^{-0.5}$ cm$^3$ s$^{-1}$, respectively, are used in the model.

Electronic recombination of both ionic and protonated glycine were already added with the rest of the glycine-related chemistry; ionization and protonation, with subsequent electronic recombination, are the primary destruction routes for glycine in all the models. Glycine is assumed to form in 5 \% of recombinations of protonated glycine (in line with values adopted for other species and with recent experimental determinations for protonated methanol and dimethyl ether), with a generic {\em total} recombination rate coefficient of $3 \times 10^{-7} (T/300 \ \mathrm{K})^{-0.5}$ cm$^3$ s$^{-1}$. The same rate is assumed for the recombination of ionized glycine; the ejection of a hydrogen atom is again assumed to occur for 5 \% of reactions, producing the radical HNCH$_2$COOH. This may accrete onto the grains to be directly hydrogenated, or may react with a selection of hydrogen-bearing gas-phase species, such as HCO, to produce glycine, as outlined in section 2.2 (although the influence of these routes on glycine abundance is neglible). Various other gas-phase and grain-surface destruction mechanisms are also available for the HNCH$_2$COOH radical, including photodissociation and further ion-molecule destruction processes.

Formation of both hydroxylamine (NH$_2$OH) and acetic acid (CH$_3$COOH) occurs on the dust-grains in this model, and each reaches peak gas-phase abundance at around 130 -- 140 K, following evaporation. In order to test the gas-phase production of glycine from these species, both of the following reactions are added to the network, assuming a rate constant of $10^{-11}$ cm$^3$ s$^{-1}$ for each:
\begin{mathletters}
\begin{eqnarray}
\mathrm{NH}_{2}\mathrm{OH}^{+} + \mathrm{CH}_{3}\mathrm{COOH} & \rightarrow \mathrm{NH}_{2}\mathrm{CH}_{2}\mathrm{COOH}^{+} + \mathrm{H}_{2}\mathrm{O} \\
\mathrm{NH}_{3}\mathrm{OH}^{+} + \mathrm{CH}_{3}\mathrm{COOH} & \rightarrow \mathrm{NH}_{3}\mathrm{CH}_{2}\mathrm{COOH}^{+} + \mathrm{H}_{2}\mathrm{O}
\end{eqnarray}
\end{mathletters}
Neither of these processes was included in the models already analyzed in section 3.

\subsection{Gas-phase formation results}

With the simple addition of reactions (11), the results are unchanged; the contribution of the gas-phase processes is never remotely significant compared to those on the dust-grains, in spite of the availability of both reactants at similar times/temperatures.

It is noteworthy that the abundances of NH$_2$OH acheived in the current models are around 3 orders of magnitude lower than those calculated by GWH; the reason for this is that the major formation route for NH$_2$OH on the grains is the addition of surface/ice-mantle NH$_2$ and OH radicals. The availability of both these reactants is significantly curtailed in the present model by the inclusion of new hydrogen-abstraction reactions for these species (see Table \ref{tab-act}) and by the more accurate consideration of the kinetics of grain-surface reactions that are mediated by activation energy barriers (see G\&P); each reactant is more readily converted back to H$_2$O or NH$_3$ before the OH and NH$_2$ radicals are able to meet. A recent observational search by Pulliam et al. (2012), using the NRAO 12\,m telescope, detected no NH$_2$OH toward a range of previously-observed sources, with fractional abundance upper limits of $8 \times 10^{-12}$ toward Sgr B2(N). The authors noted that emission from a region more compact than 5 arcsec would not be detectable with their instrument, while the model suggests that the NH$_2$OH would be at least as compact as typical hot-core molecules such as methyl formate or ethanol, based on evaporation temperatures. Nevertheless, the new model results for NH$_2$OH are more in line with the observational data.

However, a recent laboratory study suggests that the formation of NH$_2$OH by successive grain-surface hydrogenation of NO is extremely efficient (Congiu et al. 2012). While hydrogenation of NO and HNOH are already present in the network, the reaction of H with HNO has until now been assumed to lead to H-abstraction, with a barrier of 1500 K.

To increase the effect of the glycine-forming gas-phase reactions, a parallel branch is added to the surface/ice reaction set: H + HNO $\rightarrow$ HNOH. A barrier of 1500 K is also assumed; but at low temperatures, the reaction competes effectively with hydrogen diffusion, producing a reaction efficiency that is close to unity. As a result, large amounts of NH$_2$OH are formed, in this case during the cold collapse phase, rather than the warm-up phase. A total fractional abundance of $8 \times 10^{-7}$ is present in the ice by the end of collapse, i.e. as the starting point for the warm-up phase (see Table \ref{tab-ice}). This agrees well with the cold-cloud models presented by Congiu et al. (2012).

In order to further increase the influence of the gas-phase mechanisms, the rate coefficient of reaction (11a) is increased to $10^{-9}$ cm$^3$ s$^{-1}$, a plausible maximum value for this reaction.

The results of this optimized model are shown in Fig. \ref{fig-gas}. The large quantities of NH$_2$OH present in the dust-grain ices alters the fractional abundances of various species in the model by a small degree, although with a negligible direct effect on the production of glycine within the ice. The abundance of glycine is in fact slightly reduced, as a result of the somewhat lower quantity of acetic acid present in the ice. The destruction of gas-phase acetic acid is rather faster in this model, due to reaction (11a). However, when compared to an identical model with reaction (11a) switched off (not shown), the contribution of the gas-phase reaction to the total glycine formed in this model is around 2 \% of the total formed on the grains. Some of the glycine that is formed in the gas phase is re-accreted onto the grains, to evaporate later with the rest, while the remainder is destroyed by ion--molecule reactions. The immediate impact on gas-phase glycine abundance by the newly-included mechanisms may be seen in the small bump at around 130 K; less than $10^{-12} \ n_{\mathrm{H}}$ is produced. It may be noted that, if no glycine-production mechanisms of any kind were active in the ices, and all production thus depended on the gas-phase route, two abundance peaks would likely occur, at 130 K and 200 K, corresponding to peak production and peak evaporation, respectively.

On the the basis that this scenario represents the optimal conditions for gas-phase glycine formation to occur in a hot core (including the use of a very optimistic reaction rate), it appears that the particular gas-phase formation mechanism tested is unlikely to be significant, especially if the mechanisms for formation within the dust-grain ice mantles are in operation as suggested by the model. The plausibility of such a large hydroxylamine abundance as would be required for significant glycine production is also discussed in Section 5.3.

\section{Spectroscopic modeling}

In order to compare the model results more directly with observations of specific sources, a spectroscopic model is constructed that combines observationally-determined temperature and density profiles with the temperature-dependent chemical outputs of the model, to simulate emission from a selection of key molecules for which observational data exist toward the given source; this simulated emission is then convolved to allow direct comparison with the observed spectral data.

\subsection{Method}

The warm-up phase of the chemical model produces molecular fractional abundances for pre-defined output times, each corresponding to a temperature. The resolution of these output times is chosen such that the associated temperature values are separated by less than 1\%, over the 8 -- 400 K range. \citet{vdTak2000} provide such physical profiles for the envelopes of 14 nearby high-mass protostellar sources. Using these temperature--radius and density--radius relations, a radius and density is assigned to each temperature value in the chemical model outputs. The fractional abundance of each species in the model at each temperature/radius value is multiplied by its local density, to produce an absolute abundance value measured in cm$^{-3}$. The resulting molecular number-density and temperature profiles are then used to calculate the emission and absorption coefficients at each spherically-symmetric radial position in the core.

Partition functions and emission-line data from the JPL line list and the Cologne Database for Molecular Spectroscopy are imported into the code as needed, using the `Splatalogue' online database.\footnotemark \ The spectral coefficients may be calculated by this method to incorporate all molecular emission at any given frequency, allowing line blends and line confusion to be taken into account. Gaussian line profiles are assumed, using observationally-determined line widths. A channel width (frequency-bin size) much smaller than the line width is chosen (in general, 0.1 MHz or $\sim$0.08 km/s), to ensure that the lines are well resolved.

\footnotetext{splatalogue.net}

Under conditions of local thermodynamic equilibrium, the equation of radiative transfer may be parameterized thus (Rohlfs \& Wilson 2006):

\begin{equation}
I_{\nu}(s) = I_{\nu}(0) \ e^{-\tau_{\nu}(s)} \ + \ \int_{0}^{\tau_{\nu}(s)} B_{\nu}(T(\tau)) \ e^{-\tau_{\nu}} \ \mathrm{d} \tau
\end{equation}

\noindent where $I_{\nu}$ is the specific intensity, $\tau_{\nu}$ is the optical depth at frequency $\nu$, $T$ is temperature, $s$ is the line-of-sight distance, and $B_{\nu}(T)$ is the Planck function (equal to $\epsilon_{\nu}$/$\kappa_{\nu}$, under LTE). Using the gridded, spherically-symmetric input data, this equation is integrated numerically along lines of sight (using Simpson's rule), beginning directly on-source, and moving outward through parallel lines of sight until a user-defined maximum offset is reached. Line-of-sight integrals of the absorption coefficient similarly yield the optical depths. This procedure results in an intensity map of the hot core for each frequency channel.

The final step is to convolve the emission with a Gaussian beam centered on the source, using a beam-size appropriate to a chosen instrument observing at a given frequency. The intensity map is numerically integrated with the Gaussian function, over the radius of the hot-core envelope, out to a few beam widths. The technique automatically accounts for beam-dilution, and makes use of the precise emission structure to calculate an intensity value at each frequency. The resulting simulated spectrum may be directly compared with main-beam temperature ($T_{\mathrm{mb}}$) spectra obtained from the telescope for which the convolution has been produced.

\subsection{Spectroscopic simulations of NGC 6334 IRS1}

The molecular mm/sub-mm line emission from a selection of the sources observed by \citet{vdTak2000} was surveyed by \citet{Bisschop} using the James Clerk Maxwell and IRAM 30\,m telescopes, providing information on dozens on complex organic species. A member of this subset, the hot-core source NGC 6334 IRS1, is relatively closeby (1.7 kpc), and exhibits relatively narrow ($\Delta V = 5$ km/s), strong emission lines (see van der Tak et al.  2000 and Bisschop et al. 2007, and references therein). It is also well situated in the sky (Dec. (1950) = -35$^{\circ}$44') for future observation with the ALMA telescope (latitude -23$^{\circ}$). For these reasons, it may be identified as a plausible candidate for the future detection of glycine. The spectroscopic model described above is therefore used to assess the potential for the detection of glycine in this source, on the assumption that its formation and subsequent behavior are well described by the chemical model. The physical profiles provided by van der Tak et al. (2000) are used. In the case of glycine, only emission lines whose frequencies are within the observable ranges of the JCMT, IRAM\,30\,m and ALMA telescopes are modeled.

Fig. \ref{fig-specs} shows -- for illustrative purposes -- the predicted emission from this source over the 241.350 -- 241.425 GHz range, using a channel spacing of 0.488 MHz. These spectral calculations consider emission from all species in the model for which spectroscopic data exist, using only ground-state transitions. Glycine emission, at 241.373 GHz, is highlighted in red; glycine conformer I is simulated, but not conformer II, due to uncertainty in the partition functions. The upper panel of the figure shows the {\em unconvolved} emission for an on-source line of sight -- the spectrum may be considered as that expected from a pencil beam of infinite resolution directed at the source. Individual lines are identified in the upper panel; an identification is made where more than half of the emission within a channel derives from a single molecule. The emission from glycine is seen to be slightly blended with emission from glycolaldehyde (CH$_2$(OH)CHO) in the line wings, but nevertheless shows a clear peak. The middle panel shows the spectrum after convolution of the emission from the entire source (same temperature scale) using a beamwidth of 0.4 arcsec, which encompasses the strongest predicted emission region for glycine (see below) and which should be achievable with ALMA at these frequencies (although the simulation is for a single-dish telescope, rather than an array). The glycine peak is preserved, with a line strength on the order of 1 K. The lower panel of the figure shows the predicted spectrum (different temperature scale) using a beamwidth appropriate to the James Clerk Maxwell Telescope at these frequencies ($\sim$10.4 arcsec); the glycine emission line strength is on the order of 1 mK.

A selection of the strongest emission lines from methanol, methyl formate and glycine have been simulated using the method above. Methanol and methyl formate were both detected by Bisschop et al. (2007) toward this source; both are typically-observed hot-core species. In these calculations, unlike those illustrated in Fig. \ref{fig-specs}, only the emission from an individual molecule is considered at the chosen line frequency; this enables a more straight-forward comparison with the unblended lines observed by \citet{Bisschop}, and makes the calculations less computationally intensive. Furthermore, while line identifications may be made fairly easily in the unconvolved ``pencil-beam'' simulations, assessing the contributions made by each molecular line to the {\em convolved} emission spectra is much more time-consuming. The consideration of the detectability of each glycine line in the following analysis therefore does not consider potential line blends -- simply the expected line strength. More detailed assessments of the expected line emission from various species will be made in future work.

Fig. \ref{fig-profile} shows the emission profiles predicted for the line center of a strong methyl formate line (left panel) and a strong glycine line of similar frequency (right panel; glycine emission is also shown in left panel for direct comparison). The local emission strength (i.e. assuming a pencil beam) is shown in terms of brightness temperature for lines of sight targeted on radial displacements from the source position on the sky. Figure \ref{fig-map} shows the same profiles mapped onto sky coordinates; these emission maps are then convolved with the beam to produce simulated detected brightness temperatures. The most significant emission from methyl formate is seen to emanate from a region of diameter $\sim$7000 AU, or 4 arcsec. All potentially-detectable emission from glycine comes from a region $\sim$0.8 arcsec in diameter, although the most significant emission region is of diameter $\sim$0.4 arcsec, corresponding to a radius of $\sim$1000 AU.

\subsubsection{Comparison with observations}

Tables \ref{tab-rt-fast}, \ref{tab-rt-mid} and \ref{tab-rt-slow} show details of the predicted emission from each molecule, using model outputs corresponding to the {\em fast}, {\em medium} and {\em slow} warm-up models, respectively. Line frequencies are shown in GHz, along with upper-level energies in K; in some cases, each `line' represents two or more lines of identical or very similar frequencies and/or spectral characteristics. The column marked `peak simulated local intensity' indicates the maximum local emission strength (assuming a pencil beam). Beam widths represent the full-width half-maximum of the Gaussian beam used; the precise beam width appropriate to the JCMT or IRAM\,30\,m is calculated by interpolation of telescope-specific data available online. The convolved intensity of the line is the relevant quantity for discerning the detectability of a line; however, for comparison with the observed lines, the integrated intensity (in units of K km s$^{-1}$) provides a better test, as this quantity removes the dependence on the average line width ($\Delta V$); this allows all of the emission predicted by the model to be compared against all of the emission detected in the observations. The final column of these tables shows the integrated intensities measured for the methanol and methyl formate lines by Bisschop et al. (2007).

It may be seen immediately that the simulated integrated intensities of the methanol lines are an excellent match to the observed values, being generally around a factor of 2 lower, and showing only minor variation across the different warm-up timescale models. The simulations also provide a good match regardless of the upper-level energies and beam sizes (which vary according to frequency). The methyl formate emission is much more variable between model timescales; the {\em fast} model results provide an excellent match to all integrated intensities, falling within a factor less than 1.5 of the observed values. The {\em medium} timescale model produces values that are around 2 -- 3 times lower than those observed, similar to the agreement shown with the methanol lines. However, the {\em slow} warm-up timescale results produce integrated intensities only one hundredth part as large as presented in the observations of Bisschop et al. (2007). In this case, methyl formate suffers rapid gas-phase destruction, as described in Section 3.2.2, producing much lower instantaneous fractional abundances and much faster decline over time/temperature. The two faster warm-up models therefore appear the most favorable in comparison with observations.

\subsubsection{Glycine predictions}

The tables show predictions for six of the strongest glycine lines, or suites of lines. Each has an upper-level energy sufficiently low to be well-populated at the 200+ K temperatures at which glycine fractional abundances are strongest in the models. The predicted intensities between different lines (within each model) are very consistent, and peak {\em local} intensities range from around 1 K to 10 K, corresponding to {\em fast} and {\em slow} warm-up timescales, respectively. It is apparent that even for this very nearby and otherwise favorable hot-core source, the glycine emission strength is predicted to be very moderate; convolving the emission with a beam appropriate to the JCMT ($\sim$20 -- 22 arcesec) produces line strengths of no more than 10 mK in the most optimistic model. In the case of the {\em fast} warm-up model, which reproduces methyl formate emission most accurately, the glycine line strengths are no more than 250 $\mu$K.

Convolution of the emission using a beam appropriate to the IRAM\,30\,m telescope for each line ($\sim$10 -- 12 arcsec) produces line strengths of around 1, 3, and 30 mK, for {\em fast}, {\em medium}, and {\em slow} models respectively. This suggests that one of the strongest glycine emission lines could plausibly be detected toward the given source using this instrument, assuming that the {\em slow} model values are accurate; a 3-$\sigma$ line detection would require a noise level of around 10 mK (assuming no line blending, which is not modeled here). However, detection assuming either of the other model values would require impractical signal-to-noise values.

In order to make an approximate estimate of the detectability of glycine using ALMA, the same single-beam convolution method is used, along with two plausible, fixed beam sizes: a generic 1 arcsec beam, and a 0.4 arcsec beam to match the size of the region of strongest glycine emission (see Fig. \ref{fig-profile}). The reduced beam-dilution produced in these cases results in much greater line strengths, and suggests that the detection of a strong glycine line with ALMA is very plausible (line blending and model inaccuracies notwithstanding). The weakest emission predicted for any of the simulated glycine lines, with a beam width of 1 arcsec, is $\sim$87 mK, while at 0.4 arcsec the weakest line has a strength of $\sim$350 mK. Under the optimistic {\em slow} warm-up model conditions, convolved line strengths are close to 10 K for the strongest lines, using the smaller beam. For comparison, the ALMA on-line sensitivity calculator\footnotemark \ indicates that, with 32 antennae (Cycle 1) and a resolution of 0.4 arcsec, ALMA would achieve an rms of 200 mK per 1 km/s velocity bin in 1 hour.

\footnotetext{http://almascience.eso.org/call-for-proposals/sensitivity-calculator}

The line simulations therefore suggest that while detection of glycine using single-dish instruments is unlikely, even for this very favorable source, detection with ALMA is highly plausible (ignoring possible line-blending effects).

\subsection{NH$_2$OH toward W3 IRS5}

Pulliam et al. (2012) recently looked for hydroxylamine emission toward a selection of sources including Sgr B2(N), Orion KL and W3 IRS5. They found no NH$_2$OH emission using data obtained with the NRAO\,12\,m telescope, with noise levels of a few mK. Both the GWH model and the enhanced NH$_2$OH model of section 4.1 exhibit large fractional abundances for this molecule, albeit with different formation mechanisms in each case. In order to test whether the model values are consistent with the non-detections of Pulliam et al., the spectroscopic model is again applied to the chemical model results. Out of the sources observed by both Pulliam et al. (2012) and van der Tak et al. (2000), the only source present in both datasets is W3 IRS5. For this simulation, a generic line width of 9 km/s is assumed, following Pulliam et al. (2012).

Tables \ref{tab-nh2oh-1} and \ref{tab-nh2oh-2} show the results of a selection of NH$_2$OH emission-line simulations for W3 IRS5, using, respectively, the standard {\em medium} warm-up timescale model results, and the results of the model of Sec. 4.1 that employs a gas-phase glycine formation mechanism and a large initial NH$_2$OH abundance on the grains, produced by the low-temperature formation of NH$_2$OH on dust grains during the collapse phase. The results shown in Table \ref{tab-nh2oh-1} are all consistent with the non-detection; Pulliam et al. (2012) find a 1-$\sigma$ noise level of 4.3 mK in their W3 IRS5 data, while the simulated intensities are of order 10 $\mu$K or less. However, the model with enhanced NH$_2$OH formation produces maximum convolved line intensities of around 50 mK, or around 1 order of magnitude greater than the noise level. 

Since the simulated line strengths scale well with the peak fractional abundance of NH$_2$OH obtained in the chemical models, the observed values would require a peak gas-phase abundance of $<10^{-7}$ for the models to reproduce the non-detection of hydroxylamine, assuming all other parameters in the models are correct. The implication for the models is that the cold-stage conversion of NO/HNO to NH$_2$OH on the grains is too efficient, nominally by a factor of 10. This would suggest that the gas-phase formation of significant quantities of glycine is also less probable, in all sources. The apparent over-estimate of hydroxylamine production could be explained by several possibilities; one is that the chemical network does not include a sufficent number of competing reaction mechanisms for all of the states of hydrogenation between NO and NH$_2$OH. Another possibility is that the determination that the reaction H + HNO $\rightarrow$ H$_2$NO has an insignificant barrier, by Congiu et al. (2012), is dependent on experimental conditions that are inappropriate to these models, such as the ice composition, which is predominantly water in the ISM. Alternatively, the hydroxylamine abundance in the ice may be adversely affected either by photo-destruction or hydrogen abstraction processes within the ice, whose rates or barriers may be inaccurately quantified in the models.

\section{Discussion}

\subsection{Dust-grain chemistry}

The general behavior of complex molecules as simulated in the new models is broadly similar to that seen in the previous models, such as G\&H and GWH; evaporation of grain-surface molecules into the gas-phase occurs around the same temperatures and in similar quantities. However, the formation temperatures of molecules on the grains are somewhat different, due both to the distinction between surface and ice-mantle populations and to the difference in diffusion rates in each medium; HCO-related species are formed strongly around 25 K, while CH$_3$-related species are formed around 20 K -- both somewhat lower than previous models, though more in line with recent experimental results such as \cite{Oberg09a}. 

The interaction between the surface and ice-mantle chemistry in the models is complex; mobile radicals may land on the ice surface from the gas phase or may diffuse out of the mantle, to react to form more complex species, which show a net diffusion back into the mantle as more volatile species move out. The migration of mantle species to the surface as temperatures increase tends to mitigate the `trapping' effect that would be expected using a model in which there were no surface--mantle diffusion. Much complex chemistry also occurs within the mantles themselves, particularly at elevated temperatures. The importance of mantle versus surface processes is dependent on the source of the reactants, the mobilities of the reactants, the temperature, and the general composition of the surrounding ice; no {\em general} statement can therefore be made on this point other than that the diffusion within the ice mantles, and between the surface and the bulk ice, appears to play a major part in hot-core chemistry. (Although the importance for specific, individual molecules is discussed in section 3).

Aside from the effects of the differing treatment of the physical structure of the ices employed in this model, the inclusion of a more comprehensive set of hydrogen-abstraction reactions in the ice chemistry also demonstrates the key role of water and OH in producing other free radicals within the ice that may go on to form more complex organic structures. The OH radical is highly reactive, and may abstract hydrogen from stable molecules with only a small activation energy -- typically lower than the barrier to OH diffusion. Thus, hydrogen abstraction by OH may occur with a net rate equal to its rate of diffusion, in the case where the target molecule is less mobile than OH. The large quantities of water present in the ice makes OH the most commonly produced photodissociation product, caused by the weak, cosmic ray-induced, UV radiation that permeates the hot core. 

The complex molecules, such as glycolaldehyde (CH$_2$(OH)CHO), ethanol (C$_2$H$_5$OH) and methyl formate (HCOOCH$_3$), formed by the addition of HCO or CH$_3$ to the photoproducts of methanol (CH$_3$OH) are produced early, at low temperatures, before OH becomes mobile (or abundant). Their production is therefore reflective of the direct photodissociation of methanol into CH$_2$OH, CH$_3$O and CH$_3$. 

While greater temperatures increase OH mobility, below 40 K its primary destruction route (and the primary formation route for HCO) is the reaction OH + H$_2$CO $\rightarrow$ HCO + H$_2$O, which is assumed to require no activation energy. The mobility of H$_2$CO itself drives the reaction, and it is not until H$_2$CO evaporates at around 40 K that OH becomes important in providing radicals other than HCO. After this, HCO becomes less prevalent in the ices, while CH$_3$ is still produced in some abundance by methanol photodissociation, resulting in a moderate bias toward methyl-group addition over aldehyde production. At this point, ammonia (NH$_3$) and methanol, as major constituents of the ice, become the dominant reaction partners for OH, producing NH$_2$ and CH$_2$OH (CH$_3$O is also produced, but the higher activation energy for this process renders it uncompetitive versus the hydroxymethyl channel).

It is therefore at temperatures above $\sim$40 K that the formation of amino-organics becomes most active, by addition of NH$_2$ to methanol products, or to more complex radicals. However, at even higher temperatures, NH$_2$ also becomes an important source of radicals in itself, as the barriers to hydrogen abstraction by NH$_2$ are relatively low compared to its diffusion barrier, as is the case for OH. Above around 55 K, the photodissociation of water into OH typically results in the subsequent conversion of ammonia to NH$_2$, which becomes the ultimate agent of hydrogen abstraction from organic molecules in the ice. This continues until ammonia, the source of NH$_2$ radicals, evaporates at $\sim$120 K.

\subsection{Glycine formation}

The inclusion in the model of ice surface/mantle formation mechanisms for glycine indicates that it may be formed in a number of ways, whose degree of influence depends on the timescale of the hot-core evolution. The abundance of acetic acid (CH$_3$COOH) is perhaps most important in determining which mechanism is most important. In the {\em slow} warm-up model, acetic acid abundance is relatively high within the ices, and the abstraction of hydrogen by NH$_2$, followed by the addition of another amine group, is the most important glycine-formation route. The same is true, albeit to a lesser extent, for the {\em medium} warm-up timescale model. The {\em fast} warm-up model produces less acetic acid, and the addition of CH$_2$NH$_2$ and HOCO -- products of OH-induced hydrogen abstraction from methylamine and formic acid, respectively -- becomes the major formation route in conjunction with OH addition to the NH$_2$CH$_2$CO radical, which is formed by abstraction from the amino aldehyde glycinal (NH$_2$CH$_2$CHO). It is questionable as to whether the high acetic acid abundances produced by the {\em slow} model are supported by observations, indicating both that glycine abundances may be lower than suggested by that model, and that no single ice surface/mantle process of the three that are tested in this work should be strongly dominant.

It may also be noted that, in the case of the CH$_2$NH$_2$ + HOCO glycine-formation route, a plausible alternative channel, producing CH$_3$NH$_2$ + CO$_2$, may also be important. In this model a branching ratio of 1:1 is assumed; however, the accuracy of this assumption cannot be ascertained without experimental or computational data for the solid-phase process (as opposed to the gas-phase equivalent, in which glycine production would be prohibitively disfavored). The effect of a lower efficiency for glycine production through this method would be strongest in the {\em fast} warm-up scenario, although the availability of alternative routes means that glycine abundance is unlikely to fall by more than a factor of 2 as a result; conversely, a more favorable branching ratio could increase the glycine production by a similar factor under {\em fast} warm-up conditions.

A gas-phase mechanism for the formation of glycine was also tested in this model, consisting of the addition of protonated hydroxylamine (NH$_3$OH$^+$) and acetic acid (CH$_3$COOH), followed by electronic recombination. Even assuming the highest plausible rate, with large quantities of hydroxylamine and acetic acid produced on the grains, the amount of glycine produced is only on the order of $10^{-12}$$n_{\mathrm{H}}$, or 1\% of the amount formed on the dust grains. The limited comparison of simulated NH$_2$OH emission with observations suggests that the actual fractional abundance may indeed be lower than assumed in that model. It is possible that the fraction of dissociative recombination of protonated glycine that produce glycine itself may be somewhat higher than assumed (5\%), but it is unlikely to be underestimated by more than a factor of a few. The combination of favorable conditions and rates that would be required therefore suggest that gas-phase production by this method is unlikely to produce significant quantities of glycine.

The possibility, demonstrated in the models, for significant glycine production to occur through several distinct chemical routes suggests that the formation of glycine within the ice mantles is no less plausible than the formation of more commonly observed complex organic molecules.

\subsection{Detectibility of glycine}

The peak gas-phase fractional abundances for glycine of $8 \times 10^{-11}$ -- $8 \times 10^{-9}$ produced by the chemical models suggest that this molecule could be reasonably abundant in regions within hot cores. However, the LTE radiative transfer and convolution models suggest that the detection of glycine toward a favorable source may be impossible with single-dish instruments, assuming any but the most optimistic of the chemical models; this is due largely to the extremely compact emission exhibited by glycine. The model findings are therefore in agreement with the current lack of a detection of this molecule. However, it appears that the new ALMA telescope will have sufficient sensitivity and spatial resolution to easily detect the strongest glycine emission lines, assuming that the emission is not obscured by that from other molecules, and that the source is sufficiently closeby that the highly compact glycine-emission region is well resolved. Detection will likely require the careful choice of sources that are both nearby and that exhibit narrow emission lines. For this reason, detection toward the well-observed Galactic Center source Sgr B2(N) may be questionable. While ALMA should allow the distinction of different structures within that source, which contribute to the considerable line widths (Belloche et al. 2008), it is not clear whether a predicted compact glycine-emission region (of temperature $\simeq$200 K) could be resolved sufficiently well both to produce high enough line strengths for detection and to filter out contamination from other molecules. No simulations -- such as those produced above for other sources -- are possible, due to the absence of {\em both} a temperature and density profile for Sgr B2(N) in the literature. 

The most important factor in the detectibility of glycine toward hot cores is likely to be its binding energy, which determines its evaporation temperature, and thus its radius of peak fractional abundance. The simulations of glycine emission in NGC 6334 IRS1 suggest that the strongest glycine emission should emanate from a region $\sim$1000 AU in diameter within that source. This putative emission is predicted to be detectable with ALMA. The single-dish convolution of the simulated emission shown in Section 5.2, using a beam width of 0.4 arcsec as appropriate to ALMA, represents a first approximation to the detectibility of glycine at that resolution; the use of an interferometer would produce spatial filtering such that the more extended emission of other hot-core molecules could be resolved out, lessening the impact of contamination from other lines on the glycine emission.

\subsection{Combined chemical/spectroscopic models}

The combined chemical and spectroscopic models produce a very good match with observed methanol line emission toward NGC 6334 IRS1. The match for methyl formate is also good, in the case of the two shorter-timescale models; warm-up timescales of around 1 Myr may not therefore be appropriate to hot cores, although the precise binding energy of methyl formate is important to this determination. The methyl formate results also indicate that, while other mechanisms may plausibly exist for the formation of methyl formate in hot cores, the addition of methanol photo-fragments within the dust-grain ice mantles, as modeled here, appears adequate to explain observed abundances. Meanwhile, the structural isomer of methyl formate, glycolaldehyde, is almost certainly over-produced, although this may be remedied by a more careful consideration of the product branching ratios of the reaction between HCO and CH$_2$OH radicals in the ices. Branching ratios favoring an alternative [CO + CH$_3$OH] channel would seem plausible.

All of the radiative transfer calculations presented here are highly dependent on the observationally-determined temperature--density--radius profiles provided by van der Tak et al. (2000). While the simulations of methanol and methyl formate emission agree remarkably well with recent observational data, the uncertainty in the profiles naturally increases at smaller radii, and the fidelity of a spherically-symmetric model must inevitably break down where outflows or clumpy structure become important. Nevertheless, in the absence of more precise data, perhaps also to be provided by ALMA, the simple models used here provide a first quantitative estimate of the behavior and detectability of glycine in star-forming regions. A more comprehensive assessment would also incorporate the analysis of potential line blending, using spectroscopic models that include all species for which data exist, as well as an explicit treatment to simulate interferometric observations where appropriate.

\section{Conclusions}

The main conclusions of this work are summarized below:
\begin{enumerate}
\item The hot-core chemical models predict peak gas-phase glycine abundances of $\sim$$8 \times 10^{-11}$ -- $8 \times 10^{-9}$ $n_{\mathrm{H}}$ (dependent on warm-up timescale).  The peak abundance values are attained at around 200 K, when glycine evaporates from the dust grains.

\item Glycine is found to form almost exclusively within or upon dust-grain ice mantles, beginning at temperatures of $\sim$40--50 K, and ending at $\sim$120 K. The three main radical--radical addition mechanisms investigated appear to have approximately equal influence on the resultant gas-phase quantities.

\item Gas-phase glycine formation involving acetic acid (CH$_3$COOH) and protonated hydroxylamine (NH$_3$OH$^+$) is insignificant under even the most optimistic conditions.

\item Related organic species such as propionic acid, propionaldehyde (propanal), and the amino aldehyde glycinal are predicted also to attain significant gas-phase abundances, although their detectibility would be dependent also on their emission radii, via their evaporation temperatures.

\item The production of glycinal (NH$_2$CH$_2$CHO) and acetic acid (CH$_3$COOH) -- two key molecules in the formation of glycine in the ice mantles -- are strongly dependent on the earlier production of acetaldehyde (CH$_3$CHO). Acetaldehyde and acetic acid could provide observational information on potential glycine abundances in the absence of direct detection.

\item The current lack of a detection of glycine toward hot cores may be explained by its high temperature of evaporation, which results in a small emission radius. Glycine is not expected to be detectable with single-dish instruments over realizable integration times toward any hot-core source.

\item The spectroscopic model suggests glycine may be detectable using ALMA at sub-arcsecond resolutions toward bright, nearby sources with relatively narrow emission lines, such as NGC 6334 IRS1.

\item Observed methyl formate abundances are well reproduced by the two shorter-timescale models, indicating that grain-related formation mechanisms are sufficient (as suggested by previous models), and that (8--400 K) warm-up timescales of around 1 Myr may be inappropriate for hot cores. 

\item The anomalously high abundance produced by the models for the structural isomer of methyl formate, glyclolaldehyde, may be remedied by the consideration of alternative branches for the HCO + CH$_2$OH reaction in the ice mantles, favoring the formation of CO + CH$_3$OH.

\item Cosmic ray-induced photodissociation of water ice, to produce OH, is crucial to the production of radicals within the bulk ices. The OH radicals easily abstract hydrogen from other molecules, including ammonia, to produce reactive molecular radicals that may combine to form more complex species. 

\item NH$_2$, formed by hydrogen abstraction from ammonia by OH, also appears to be an important agent of radical propagation within the ice mantles. This process, and the related production of amino-organics, becomes most significant at temperatures between $\sim$40--120 K. The observed nitrogen content of hot-core molecules may thus be related to the timescale spent in this temperature range.

\item Much, if not most, of the chemistry that produces complex organic molecules occurs within the ice mantles, while the more rapid diffusion that obtains on the ice surface allows different reaction kinetics to occur concurrently. Thus, the explicit treatment of these two phases, their coupling via thermal diffusion, and the interaction of the dust/ice surface with the gas phase are all essential elements in the chemical model.

\item The simple spectroscopic model used here provides a direct comparison between chemical models and observational spectroscopic data for complex molecules in specific sources, an element that has been lacking until now in the study of hot cores.

\end{enumerate}

\acknowledgements

The author thanks the referee for providing helpful comments, and K. {\"O}berg for a critical reading of the manuscript and many valuable discussions. This work is funded by the NASA Astrophysics Theory Program, grant number NNX11AC38G.

\newpage

\begin{deluxetable}{lc}
\tabletypesize{\small}
\tablecaption{\label{tab-init} Initial fractional abundances, with respect to total hydrogen, of elements and H$_2$ at
the start of the collapse phase.} 
\tablewidth{0pt} 
\tablehead{ \colhead{Species, $i$} & \colhead{$n_{i}/n_{H}$
\tablenotemark{\dag}}} 
\startdata
H$_2$ & $0.499 $ \\
H & $2.0(-3) $ \\
He & $0.09$ \\
C & $1.4(-4)$ \\
N & $7.5(-5)$ \\
O & $3.2(-4)$ \\
S & $8.0(-8)$ \\
Na & $2.0(-8)$ \\
Mg & $7.0(-9)$ \\
Si & $8.0(-9)$ \\
P & $3.0(-9)$ \\
Cl & $4.0(-9)$ \\
Fe & $3.0(-9)$ \\
\enddata
\tablenotetext{\dag}{$A(B)=A^{B}$}
\end{deluxetable}

\newpage

\begin{deluxetable}{lclclcl}
\tabletypesize{\scriptsize}
\tablecaption{\label{tab-surf} A selection of barrierless surface/mantle radical-addition reactions relevant to glycine or related species included in the chemical network}
\tablewidth{0pt}
\tablecolumns{7}
\tablehead{
\colhead{Reaction} 
}
\startdata
 HOCO     &  +  & CH$_2$NH$_2$    &  $\rightarrow$  &    NH$_2$CH$_2$COOH   \\
                                                 &&   &  $\rightarrow$  &    CH$_3$NH$_2$     &  +  & CO$_2$   \\
 NH$_2$   &  +  & CH$_2$COOH      &  $\rightarrow$  &    NH$_2$CH$_2$COOH   \\
 OH       &  +  & NH$_2$CH$_2$CO  &  $\rightarrow$  &    NH$_2$CH$_2$COOH   \\
 H        &  +  & HNCH$_2$COOH     &  $\rightarrow$  &   NH$_2$CH$_2$COOH   \\
 CH$_2$OH &  +  & HNCH$_2$COOH   &  $\rightarrow$  &     NH$_2$CH$_2$COOH &  +  & H$_2$CO   \\
 CH$_3$O  &  +  & HNCH$_2$COOH   &  $\rightarrow$  &     NH$_2$CH$_2$COOH &  +  & H$_2$CO   \\
 HCO      &  +  & HNCH$_2$COOH    &  $\rightarrow$  &    NH$_2$CH$_2$COOH &  +  & CO   \\
\enddata
\tablecomments{Not all reactions involving each species are shown. Equal branching ratios are assumed, where multiple channels are given. The formation of the relevant radicals may occur through photodissociation or through hydrogen-abstraction reactions, detailed in Table \ref{tab-act}.}
\end{deluxetable}

\newpage

\begin{deluxetable}{rlccl}
\tabletypesize{\scriptsize}
\tablecaption{\label{tab-act} Activation energy barrier information for important barrier-mediated surface/mantle reactions}
\tablewidth{0pt}
\tablecolumns{13}
\tablehead{
\colhead{\#} & \colhead{Reaction} & \colhead{$E_{\mathrm{A}}$} & \colhead{Width} & \colhead{Ref.} \\
\colhead{} &            \colhead{} &             \colhead{(K)} &                 \colhead{(\AA)} & 
}
\startdata

 1 & H        +  CO               $\rightarrow$ HCO                          & --- & ---    & Andersson (priv. comm.) \\ 
 2 & H        +  H$_2$CO          $\rightarrow$ CH$_2$OH                     & 4500 & 1.35    & Goumans (priv. comm.) \\ 
 3 & H        +  H$_2$CO          $\rightarrow$ CH$_3$O                      & 2320 & 1.35    & Goumans (priv. comm.) \\ 
 4 & H        +  H$_2$CO          $\rightarrow$ HCO       + H$_2$            & 2960 & 1.22    & Goumans (priv. comm.) \\ 
 5 & H        +  CH$_3$OH         $\rightarrow$ CH$_2$OH  + H$_2$            & 4380 & 1.0     & Goumans (priv. comm.) \\ 
 6 & H        +  CH$_3$OH         $\rightarrow$ CH$_3$O   + H$_2$            & 6640 & 1.0     & Goumans (priv. comm.) \\ 
 7 & H        +  CH$_4$           $\rightarrow$ CH$_3$    + H$_2$            & 5940 & 2.17    & Andersson et al. (2009); Garrod \& Pauly (2011)  \\ 
 8 & H        +  CH$_3$CHO        $\rightarrow$ H$_2$     + CH$_3$CO         & 2120 & 1.0     & Warnatz (1984) \\
\enddata
\tablecomments{Reactions marked {\em Evans-Polanyi} use the parameters provided by Dean \& Bozzelli (1999) unless otherwise marked with the reference reactions used to fit the Evans-Polanyi relationship. Reactions for which insufficient data exist for an Evans-Polanyi are assigned the activation-energy value of the similar reaction indicated. Reactions marked {\em estimate} are assigned a generic value appropriate to that reaction type.}
\end{deluxetable}

\newpage

\begin{deluxetable}{lrrr}
\tabletypesize{\small}
\tablecaption{\label{tab-bind} Binding energies and diffusion barriers used for a selection of important atoms and molecules, representative of adhesion to an amorphous ice surface.}
\tablewidth{0pt} 
\tablehead{ \colhead{Species} & \colhead{$E_{\mathrm{des}}$} & \colhead{$E_{\mathrm{swap}}$} & \colhead{$E_{\mathrm{dif}}$} }
\startdata
H                   &   450     &   315  &   158    \\
H$_2$               &   430     &   301  &   150    \\
OH                  &  2850     &  1995  &   998    \\
H$_2$O              &  5700     &  3990  &  1995    \\
CH$_2$              &  1050     &   735  &   368    \\
CH$_3$              &  1175     &   823  &   411    \\
CH$_4$              &  1300     &   910  &   455    \\
NH                     &  2380     &  1666  &  833    \\
\enddata
\end{deluxetable}

\newpage

\begin{deluxetable}{lcll}
\tabletypesize{\scriptsize}
\tablecaption{\label{tab-gas} A selection of gas-phase reactions relevant to glycine included in the chemical network}
\tablewidth{0pt}
\tablecolumns{4}
\tablehead{
\colhead{Reaction} & \colhead{} & \colhead{} & \colhead{Rate}
}
\startdata
NH$_2$CH$_2$COOH          + $h\nu$ (CR)  & $\rightarrow$ &      NH$_2$CH$_2$CO  +  OH                            &   $1.50 \times 10^{3} \ \zeta_{0}$ \\
NH$_2$CH$_2$COOH          + $h\nu$ (CR)  & $\rightarrow$ &      CH$_2$NH$_2$    +  HOCO                          &   $5.00 \times 10^{2} \ \zeta_{0}$ \\
NH$_2$CH$_2$COOH          + $h\nu$ (CR)  & $\rightarrow$ &      CH$_2$COOH      +  NH$_2$                        &   $1.00 \times 10^{4} \ \zeta_{0}$ \\
NH$_2$CH$_2$COOH          + $h\nu$ (CR)  & $\rightarrow$ &      HNCH$_2$COOH    +  H                             &   $1.00 \times 10^{3} \ \zeta_{0}$ \\

NH$_2$CH$_2$COOH         +  $h\nu$                & $\rightarrow$ &      NH$_2$CH$_2$CO  +  OH                            &   $5.00 \times 10^{-10} \exp(-1.70 \times A_{V}) \ $s$^{-1} $ \\
NH$_2$CH$_2$COOH         +  $h\nu$                & $\rightarrow$ &      CH$_2$NH$_2$    +  HOCO                          &   $9.00 \times 10^{-10} \exp(-1.60 \times A_{V}) \ $s$^{-1} $ \\
NH$_2$CH$_2$COOH         +  $h\nu$                & $\rightarrow$ &      CH$_2$COOH      +  NH$_2$                        &   $1.00 \times 10^{-09} \exp(-1.70 \times A_{V}) \ $s$^{-1} $ \\
NH$_2$CH$_2$COOH         +  $h\nu$                & $\rightarrow$ &      HNCH$_2$COOH    +  H                             &   $5.00 \times 10^{-10} \exp(-1.70 \times A_{V}) \ $s$^{-1} $ \\
\enddata
\tablecomments{The cosmic ray ionization rate, $\zeta_{0}$, is set to $1.3 \times 10^{-17}$ s$^{-1}$ throughout the model. $A_{V}$ denotes visual extinction. CR denotes cosmic ray induced UV radiation.}
\tablenotetext{\dag}{These reactions and rates are used only in the models presented in Section 4.}
\end{deluxetable}

\newpage

\begin{deluxetable}{lcc}
\tabletypesize{\small}
\tablecaption{\label{tab-mod} Warm-up phase model parameters}
\tablewidth{0pt}
\tablehead{
\colhead{Model} &
\colhead{Time to reach} &
\colhead{Time to reach} \\
\colhead{} &
\colhead{200 K} &
\colhead{400 K} 
}
\startdata
Fast         &  $5 \times 10^{4}$ &  $7.12 \times 10^{4}$  \\
Medium   &   $2 \times 10^{5}$ &  $2.85 \times 10^{5}$  \\
Slow       &   $1 \times 10^{6}$ &  $1.43 \times 10^{6}$  \\
\enddata
\end{deluxetable}

\newpage

\begin{deluxetable}{lcccccc}
\tabletypesize{\small}
\tablecaption{\label{tab-ice} Selection of protostellar ice compositions from the literature, and post-collapse model values, as a percentage of H$_2$O}
\tablewidth{0pt}
\tablehead{
\colhead{Species} &
\colhead{W33A \tablenotemark{a}} &
\colhead{NGC 7538 IRS 9 \tablenotemark{b}} &
\colhead{Sgr A* \tablenotemark{b}} &
\colhead{Typical} &
\colhead{Typical} &
\colhead{Collapse-phase} \\
\colhead{} &
\colhead{} &
\colhead{} &
\colhead{} &
\colhead{low-mass \tablenotemark{c}} &
\colhead{high-mass \tablenotemark{c}} &
\colhead{model values}
}
\startdata
CO                  &  8    &  16  &  $<$12 &  29   &   13   &  57      \\
CO$_2$           &  13   &  22 &  14       &  29   &   13   &  18    \\
CH$_4$           &  1.5  &  2   &  2        &  5   &    2   &  1.9          \\
H$_2$CO         &  6    &  4   &  $<$3   &  $\leq$2   &   $\leq$2    &  1.6       \\
CH$_3$OH       &  18  &  5   &  $<$4   &  3   &   4    &  6.9      \\
NH$_3$           &  15  &  13  &  20--30 &  5   &   5    &  18       \\
NH$_2$OH       &  --  &  --  &  -- &  --   &   --    &  0 / 0.55 \tablenotemark{d}       \\

\enddata
\tablenotetext{a}{Gibb et al. (2000)}
\tablenotetext{b}{See Gibb et al. (2000) for original references}
\tablenotetext{c}{{\"O}berg et al. (2011)}
\tablenotetext{d}{Standard value / value using enhanced NH$_2$OH production described in Sec. 4.1}

\end{deluxetable}

\newpage

\begin{figure*}
\center
\includegraphics[width=0.425\textwidth]{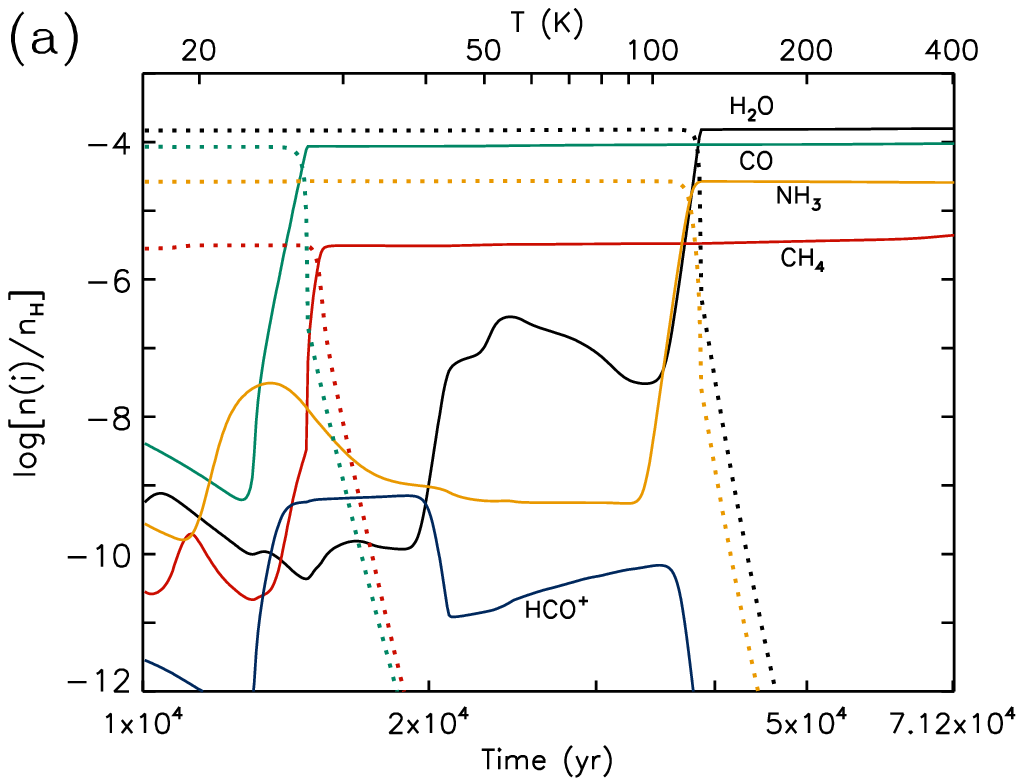}
\includegraphics[width=0.425\textwidth]{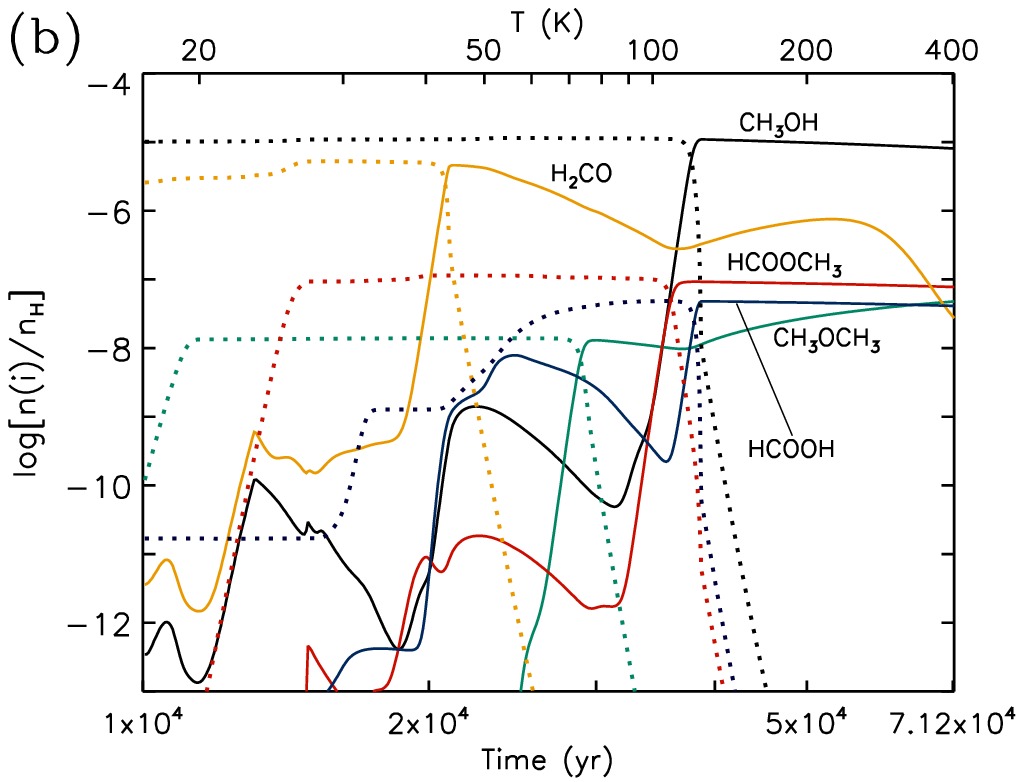}
\includegraphics[width=0.425\textwidth]{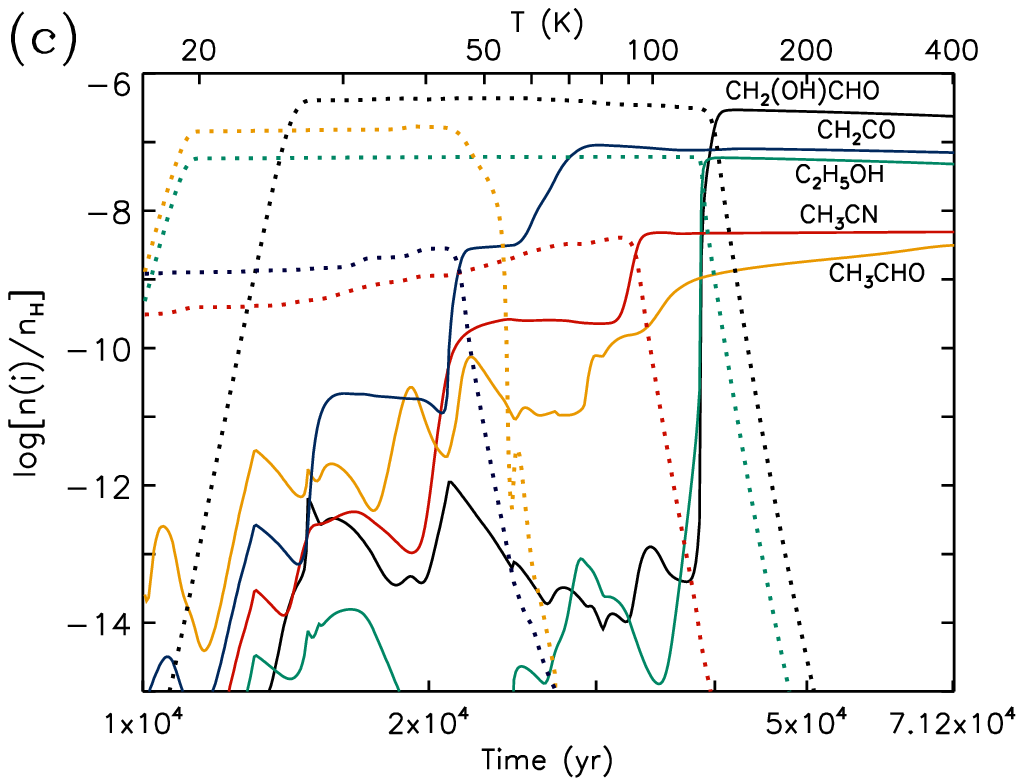}
\includegraphics[width=0.425\textwidth]{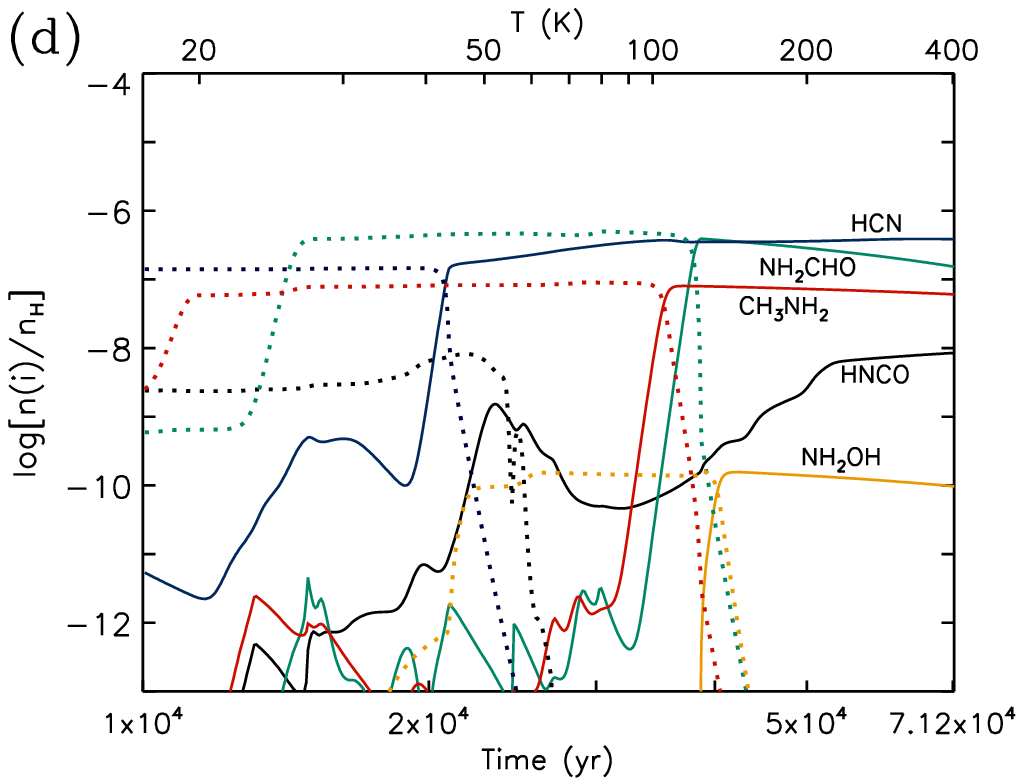}
\includegraphics[width=0.425\textwidth]{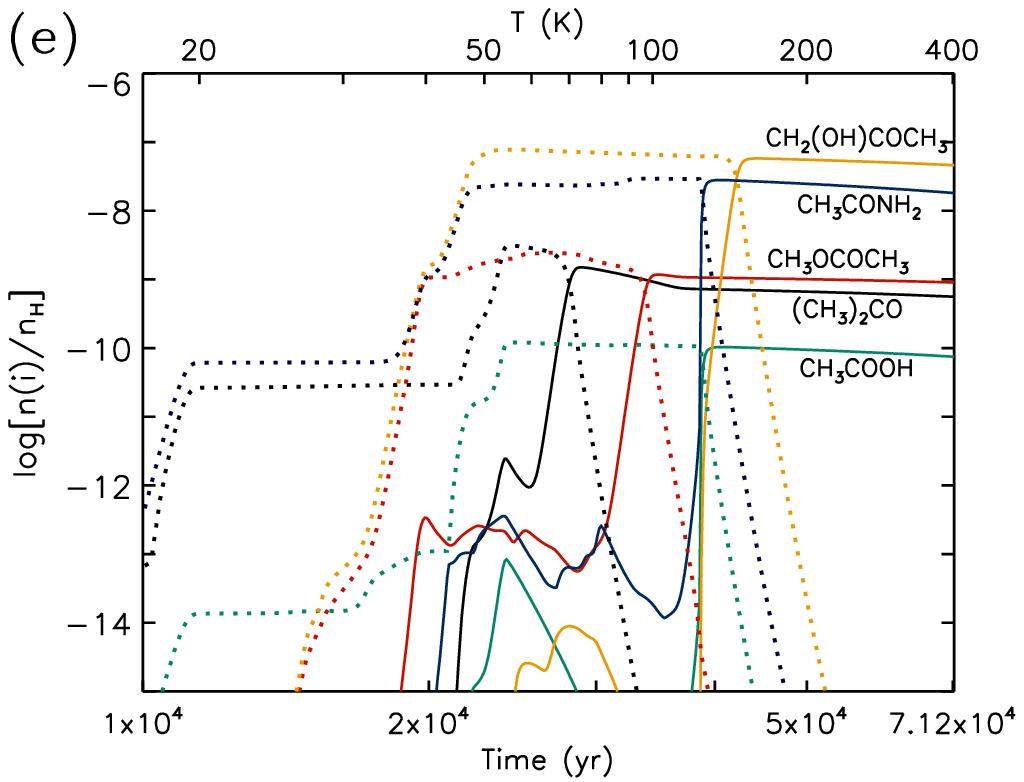}
\includegraphics[width=0.425\textwidth]{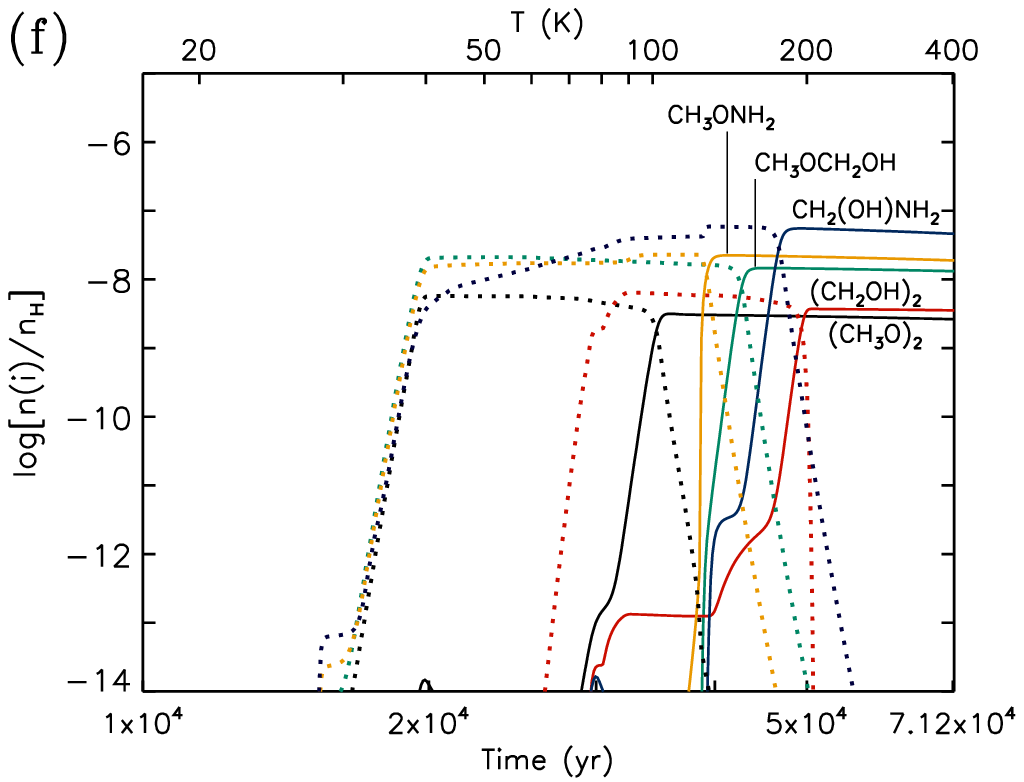}
\includegraphics[width=0.425\textwidth]{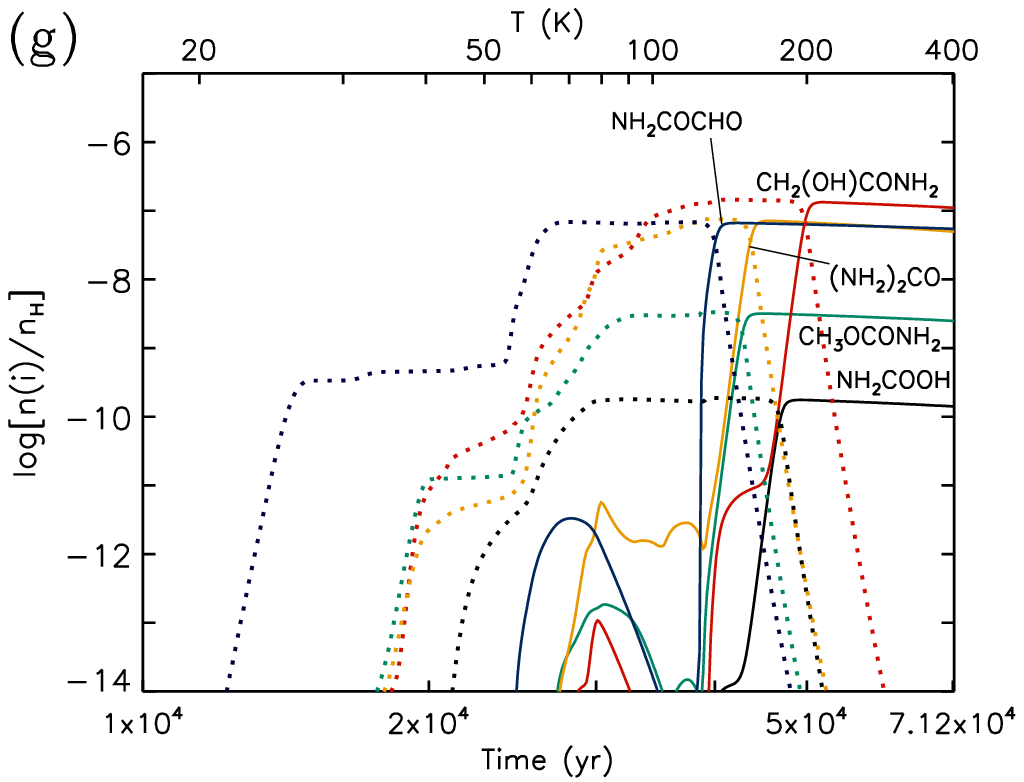}
\includegraphics[width=0.425\textwidth]{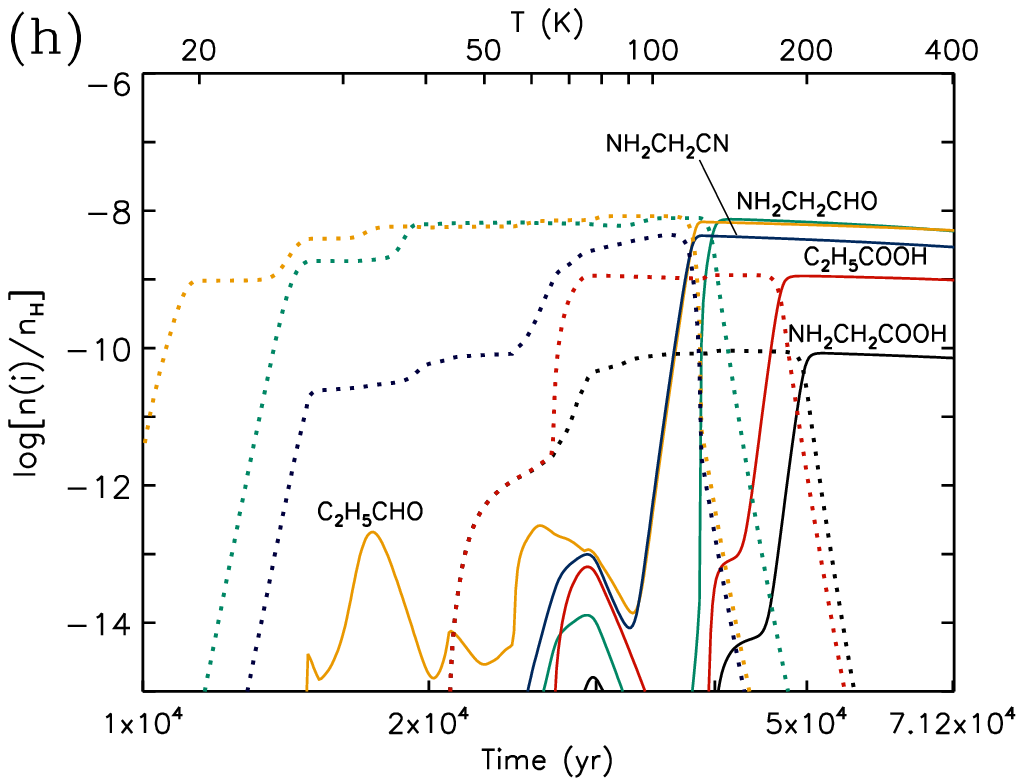}
\caption{\label{fig-fast} Time-dependent fractional abundances of a selection of chemical species, produced by the {\em \bf fast} warm-up timescale model. Solid lines indicate gas-phase species; dotted lines of the same color indicate ice-mantle (surface + bulk) abundances of the same species.}
\end{figure*}

\begin{figure*}
\center
\includegraphics[width=0.425\textwidth]{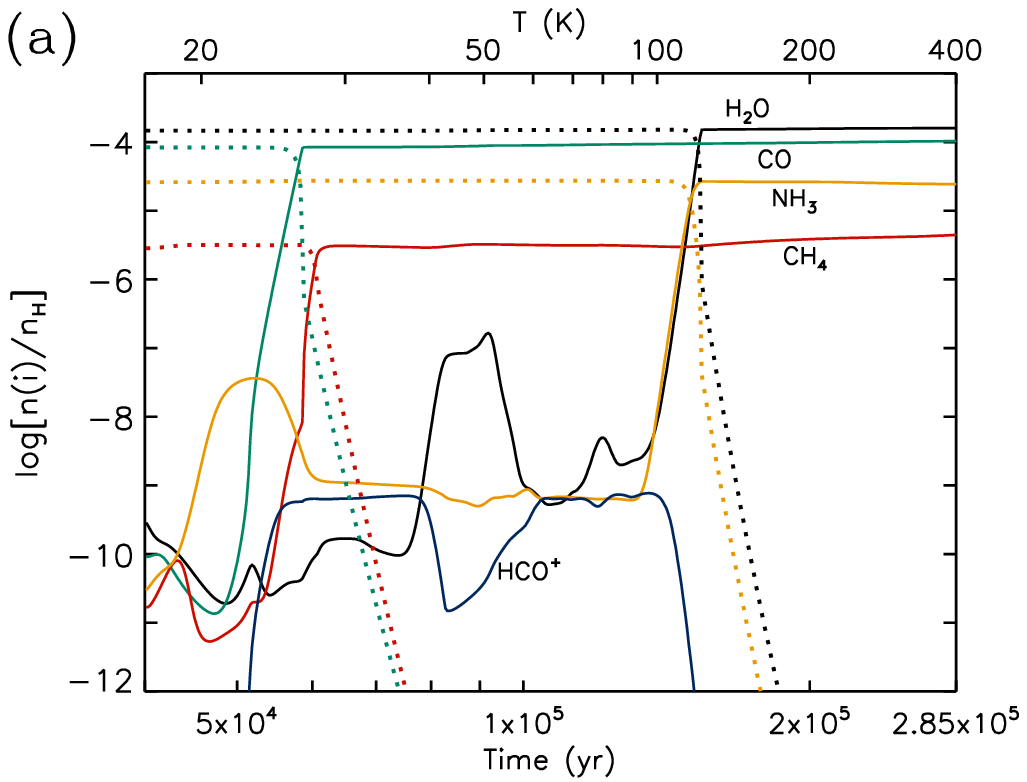}
\includegraphics[width=0.425\textwidth]{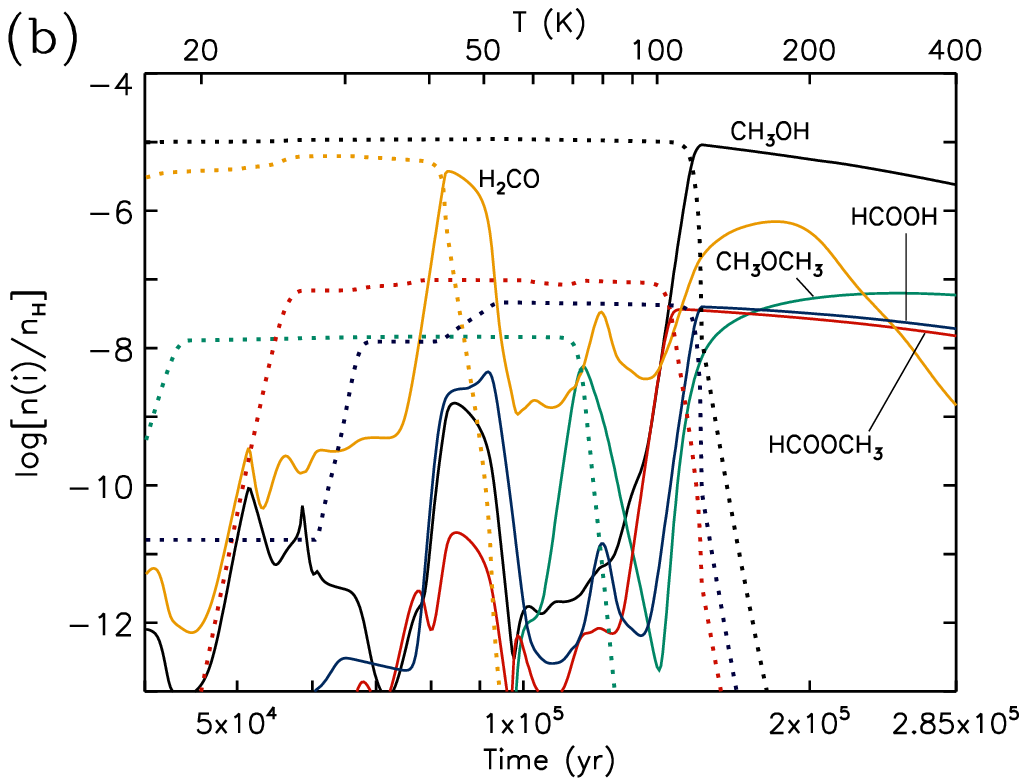}
\includegraphics[width=0.425\textwidth]{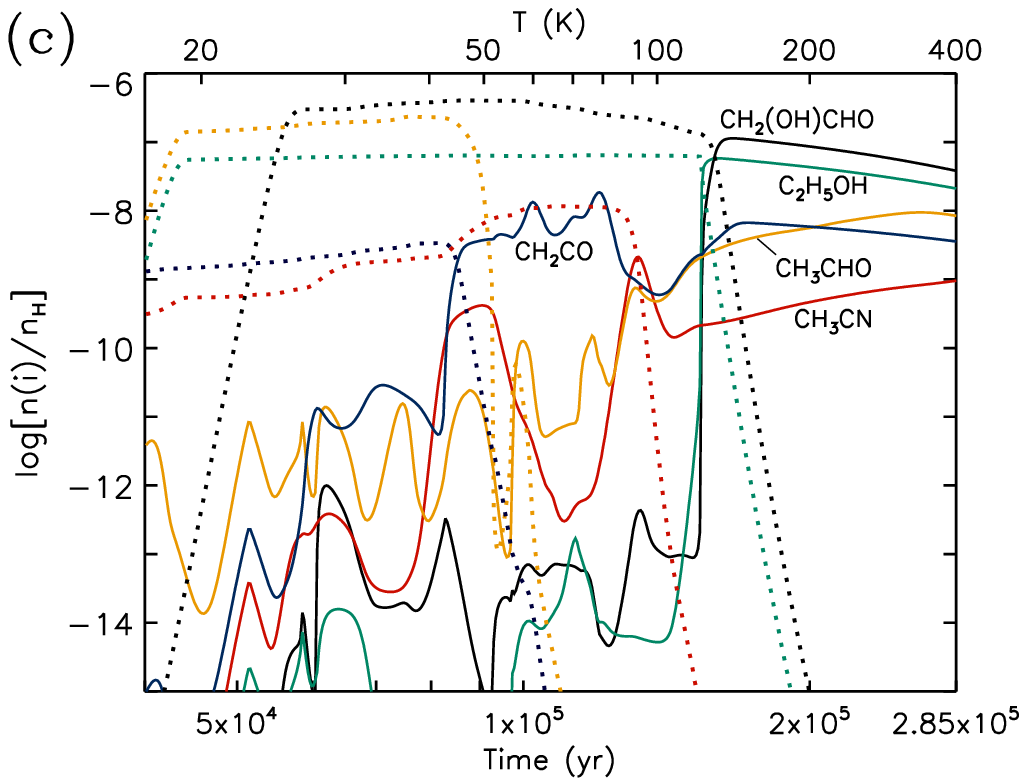}
\includegraphics[width=0.425\textwidth]{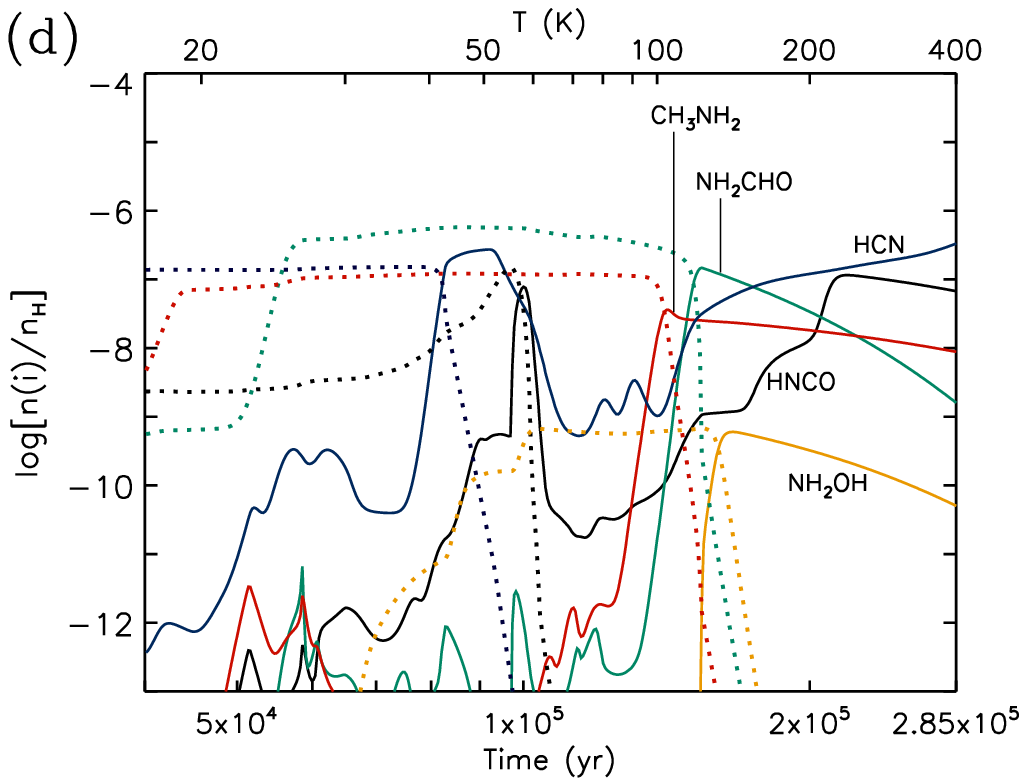}
\includegraphics[width=0.425\textwidth]{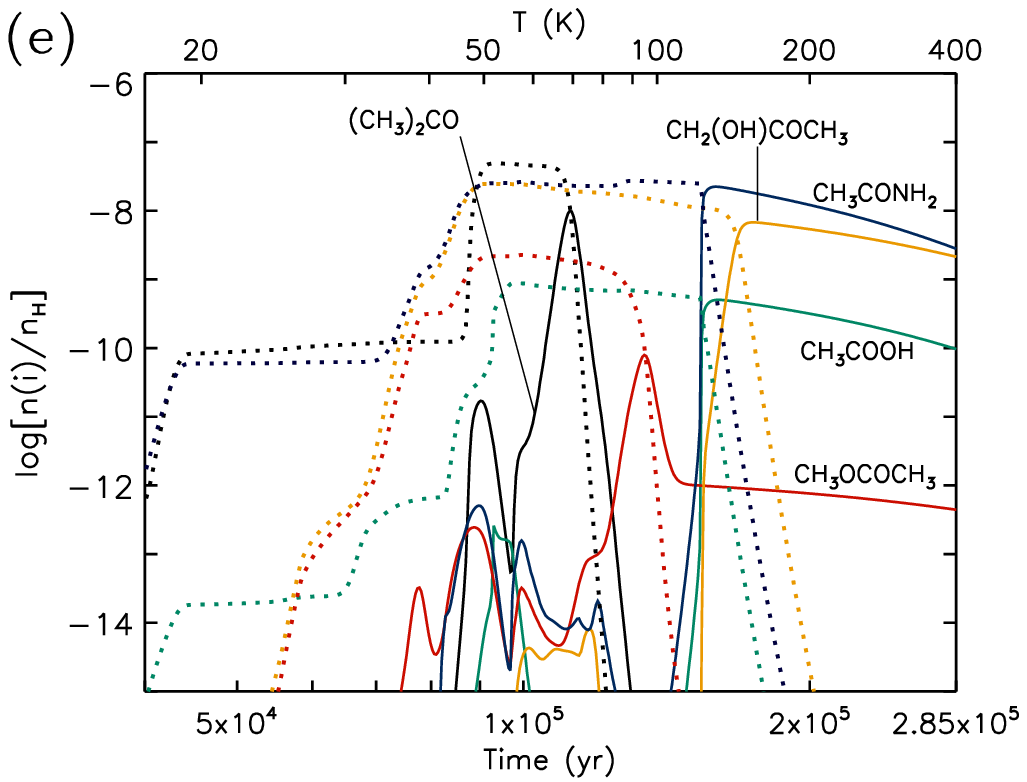}
\includegraphics[width=0.425\textwidth]{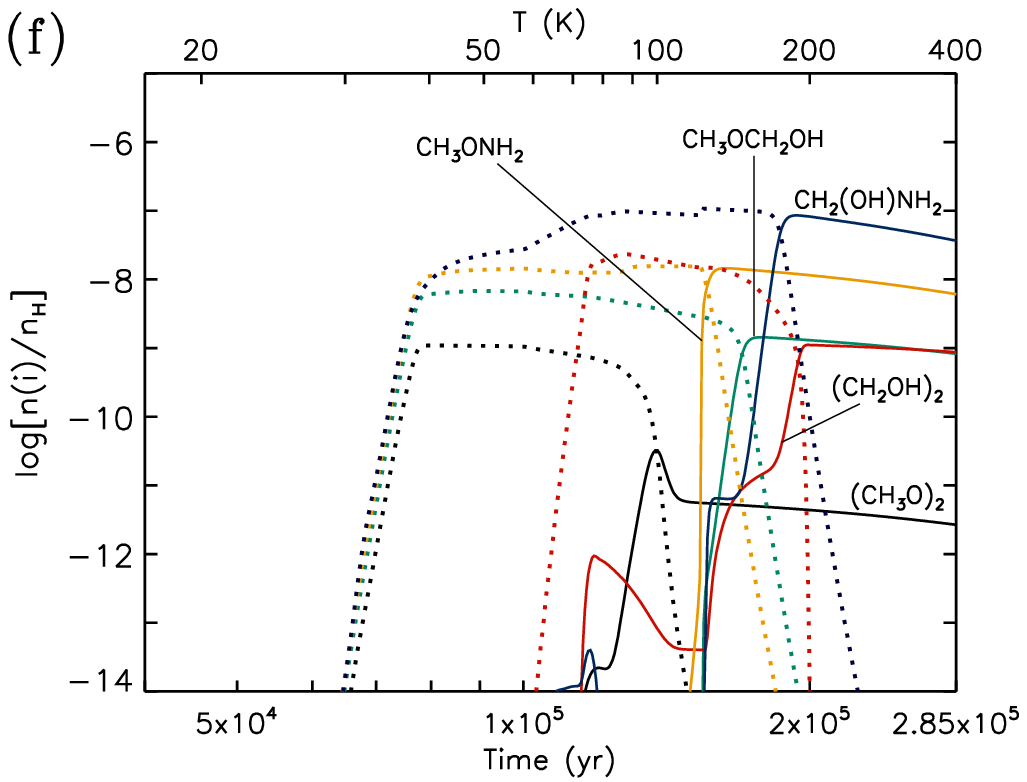}
\includegraphics[width=0.425\textwidth]{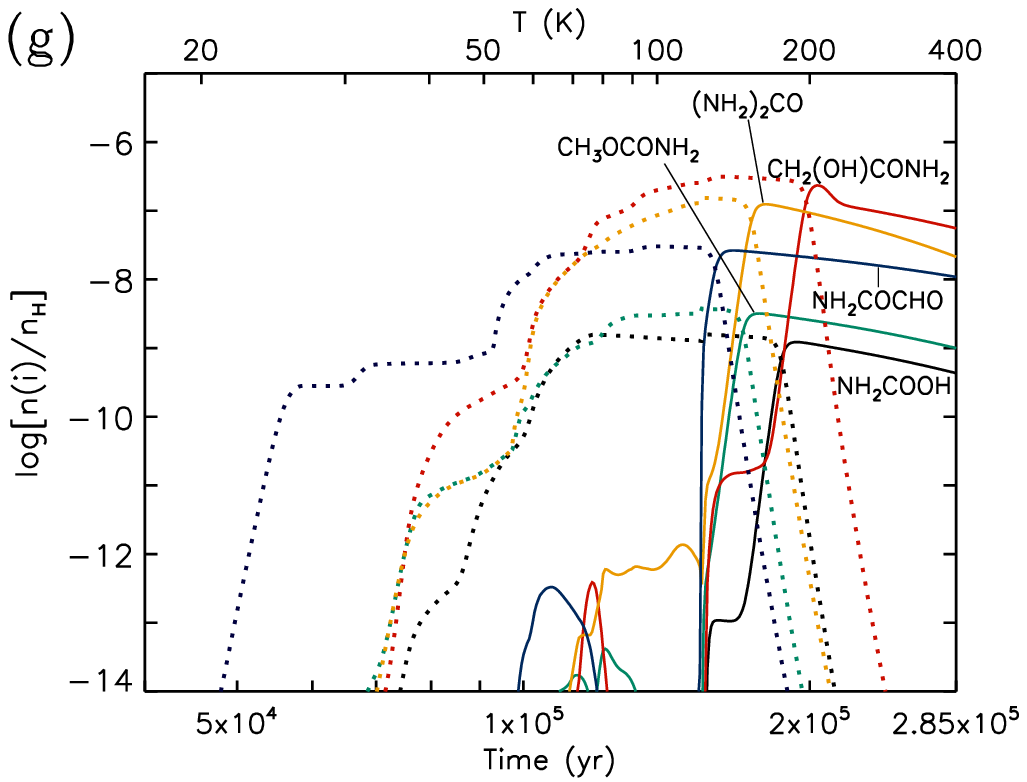}
\includegraphics[width=0.425\textwidth]{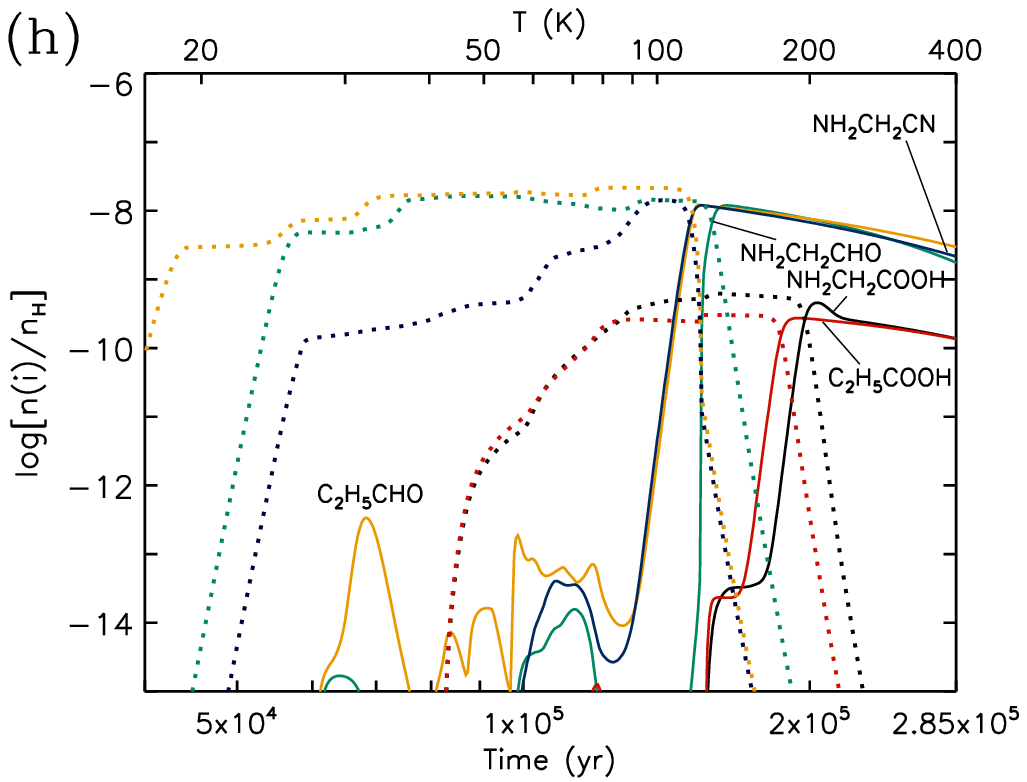}
\caption{\label{fig-mid} Time-dependent fractional abundances of a selection of chemical species, produced by the {\em \bf medium} warm-up timescale model. Solid lines indicate gas-phase species; dotted lines of the same color indicate ice-mantle (surface + bulk) abundances of the same species.}
\end{figure*}

\begin{figure*}
\center
\includegraphics[width=0.425\textwidth]{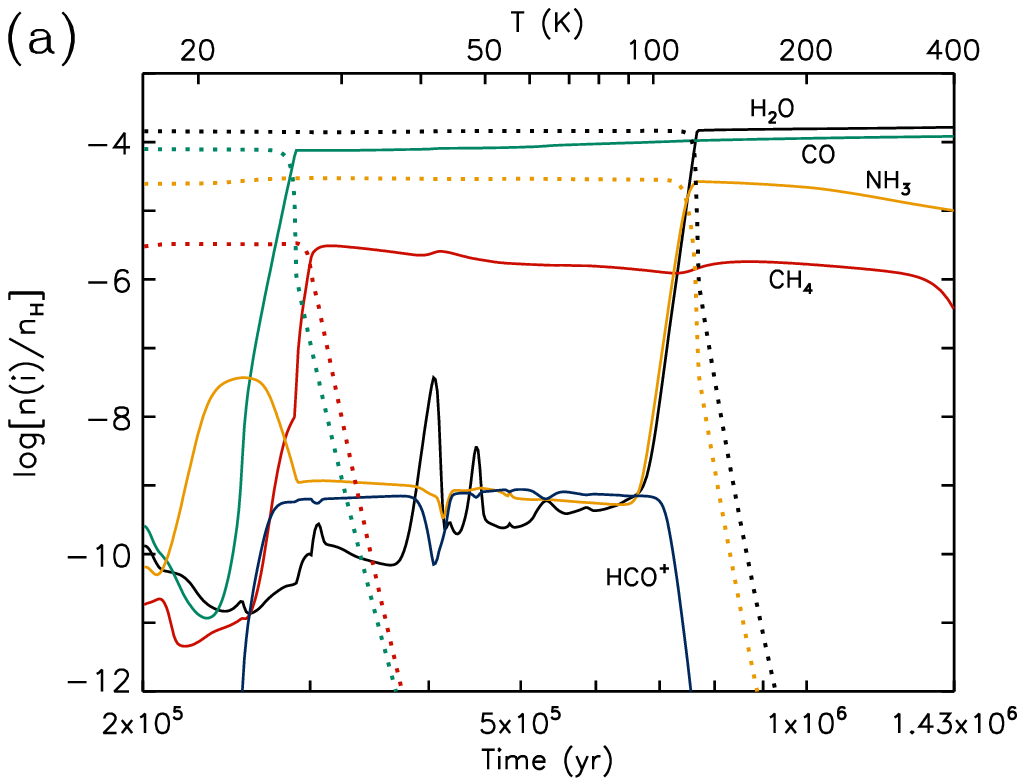}
\includegraphics[width=0.425\textwidth]{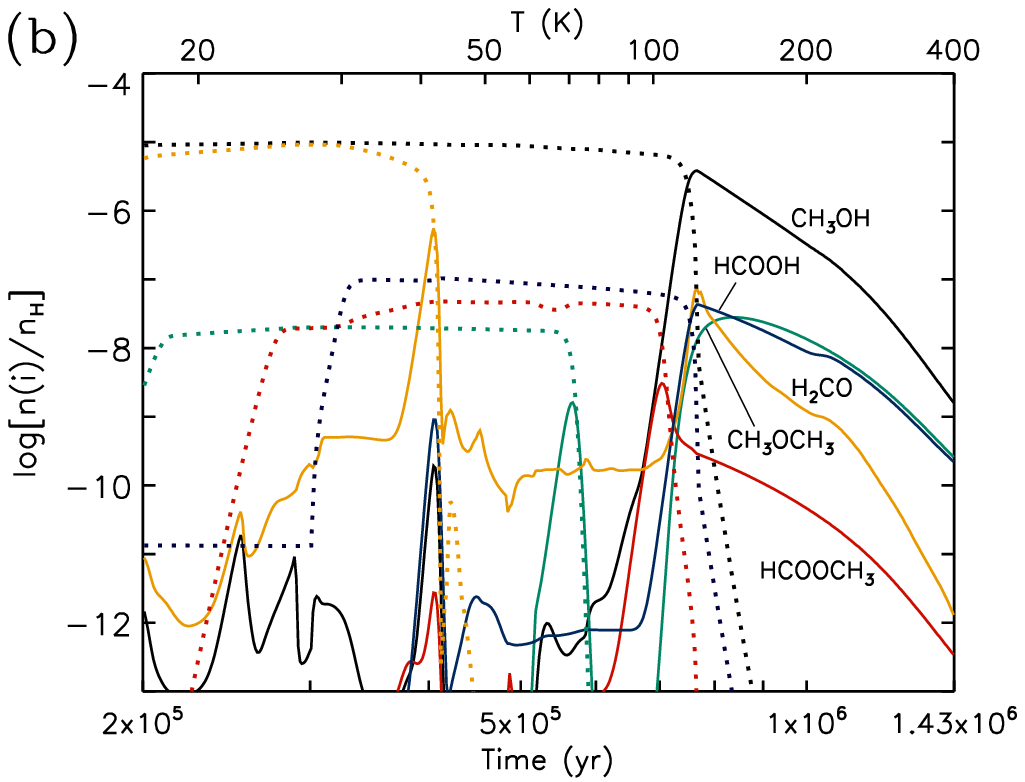}
\includegraphics[width=0.425\textwidth]{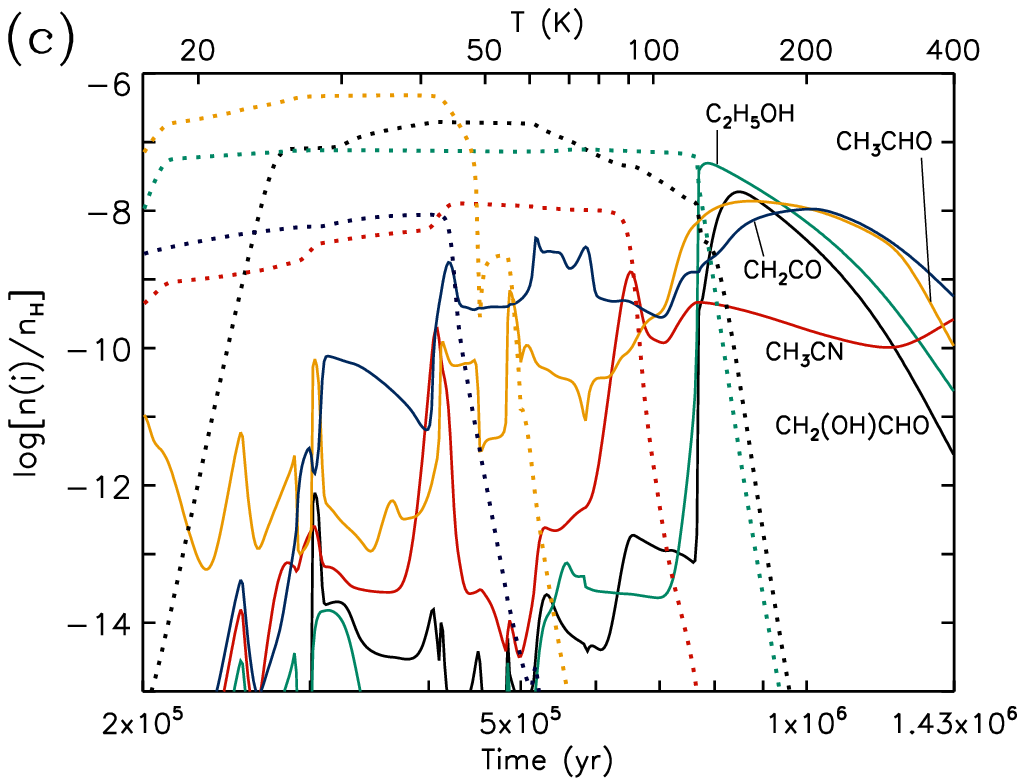}
\includegraphics[width=0.425\textwidth]{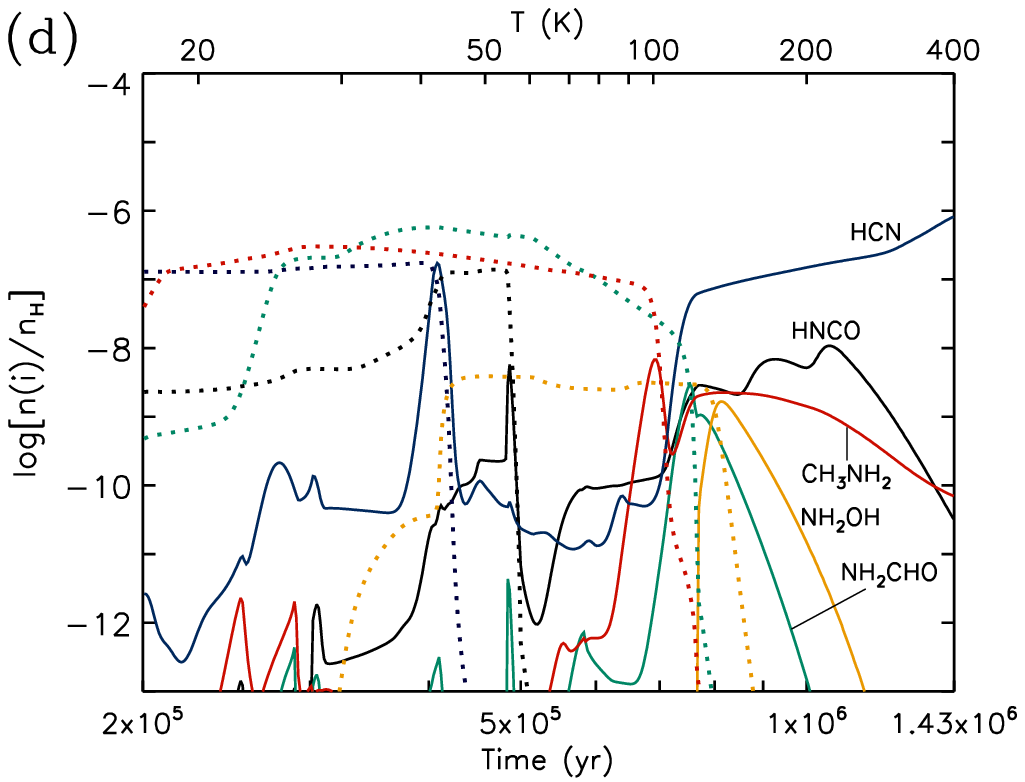}
\includegraphics[width=0.425\textwidth]{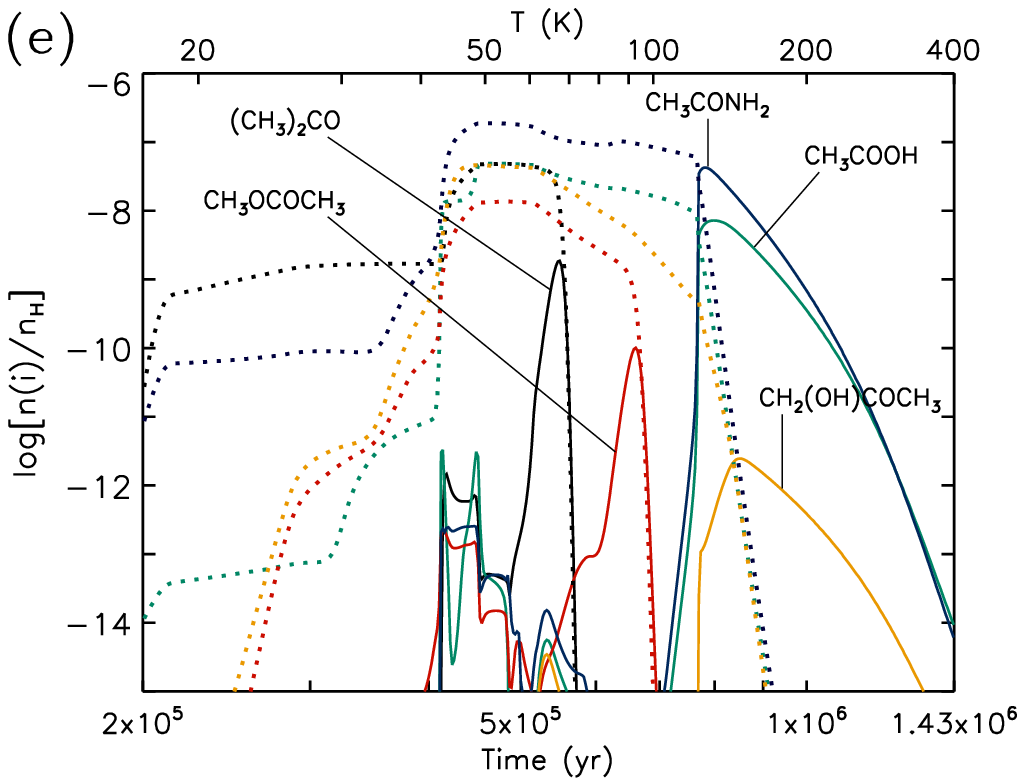}
\includegraphics[width=0.425\textwidth]{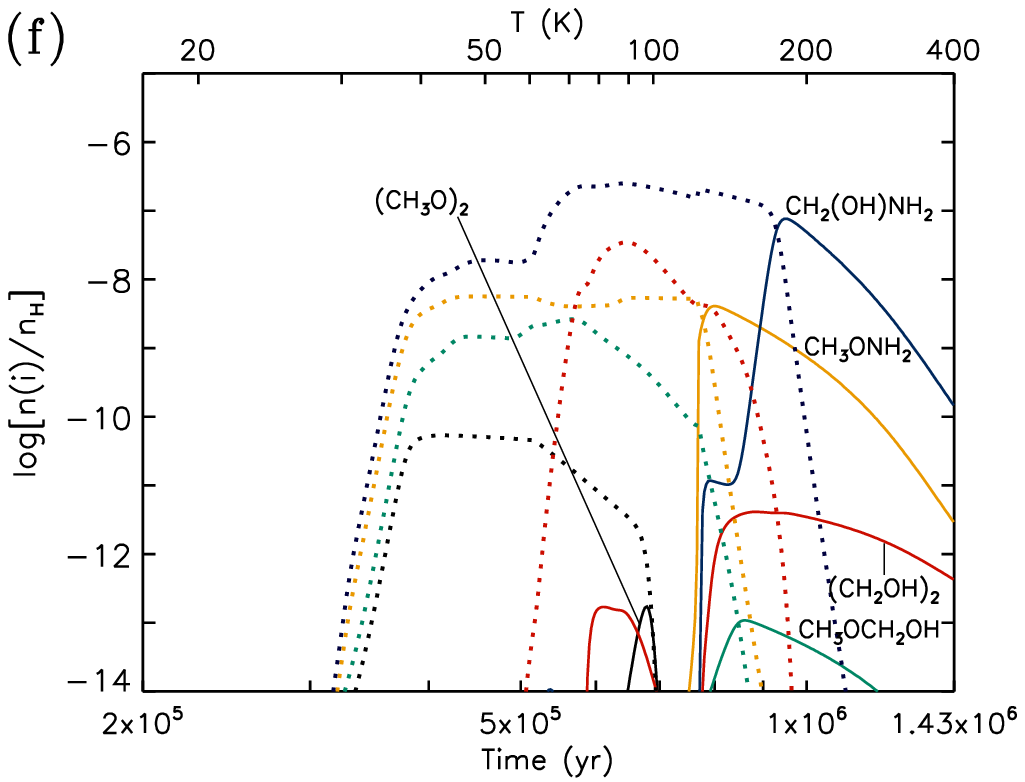}
\includegraphics[width=0.425\textwidth]{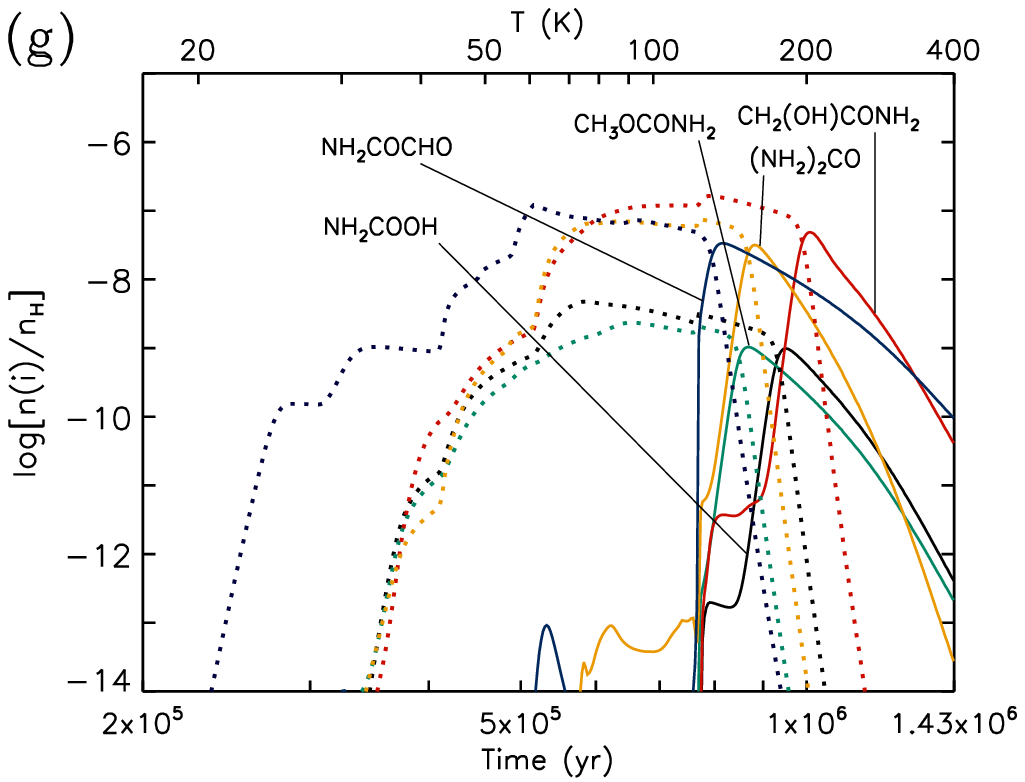}
\includegraphics[width=0.425\textwidth]{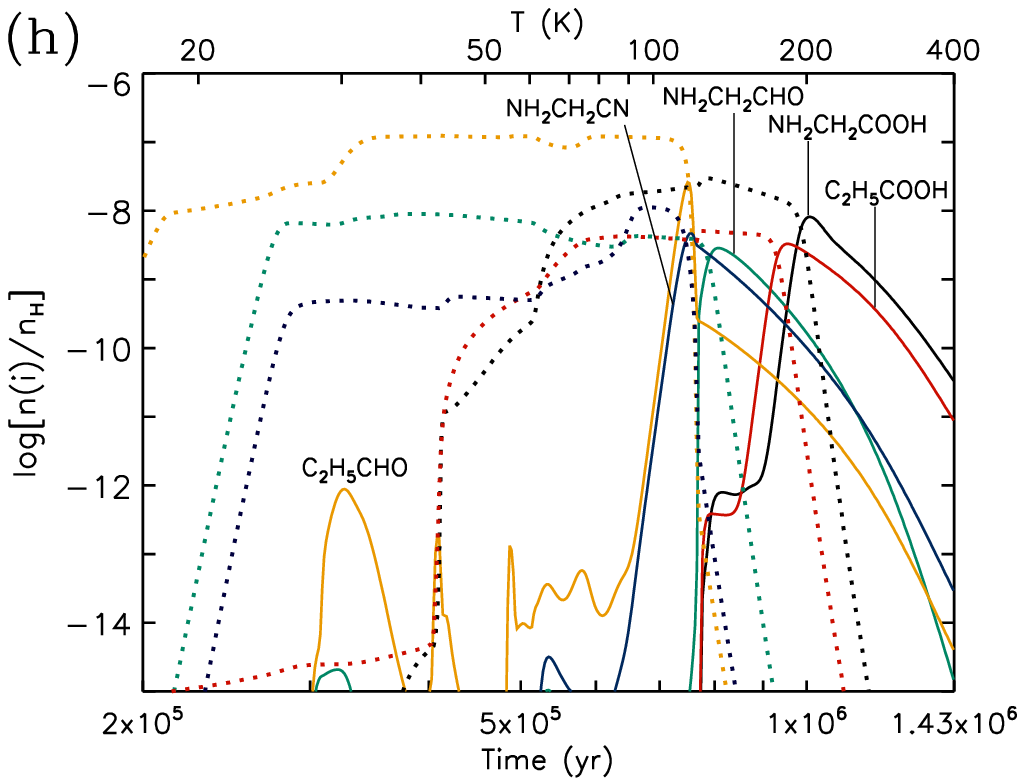}
\caption{\label{fig-slow} Time-dependent fractional abundances of a selection of chemical species, produced by the {\em \bf slow} warm-up timescale model. Solid lines indicate gas-phase species; dotted lines of the same color indicate ice-mantle (surface + bulk) abundances of the same species.}
\end{figure*}

\begin{figure*}
\center
\includegraphics[width=0.475\textwidth]{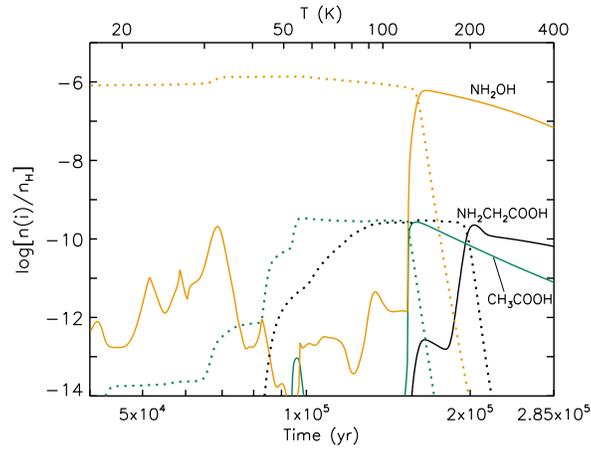}
\caption{\label{fig-gas} Time-dependent gas-phase and ice-mantle fractional abundances of hydroxylamine (NH$_2$OH), acetic acid (CH$_3$COOH) and glycine (NH$_2$CH$_2$COOH), using the {\em medium} warm-up timescale model with the addition of a fast gas-phase glycine-formation mechanism and a high initial NH$_2$OH abundance in the ice mantles, as discussed in Sec. 4.1. Solid lines indicate gas-phase species; dotted lines of the same color indicate ice-mantle (surface + bulk) abundances of the same species.}
\end{figure*}

\begin{figure*}
\includegraphics[width=1.0\textwidth]{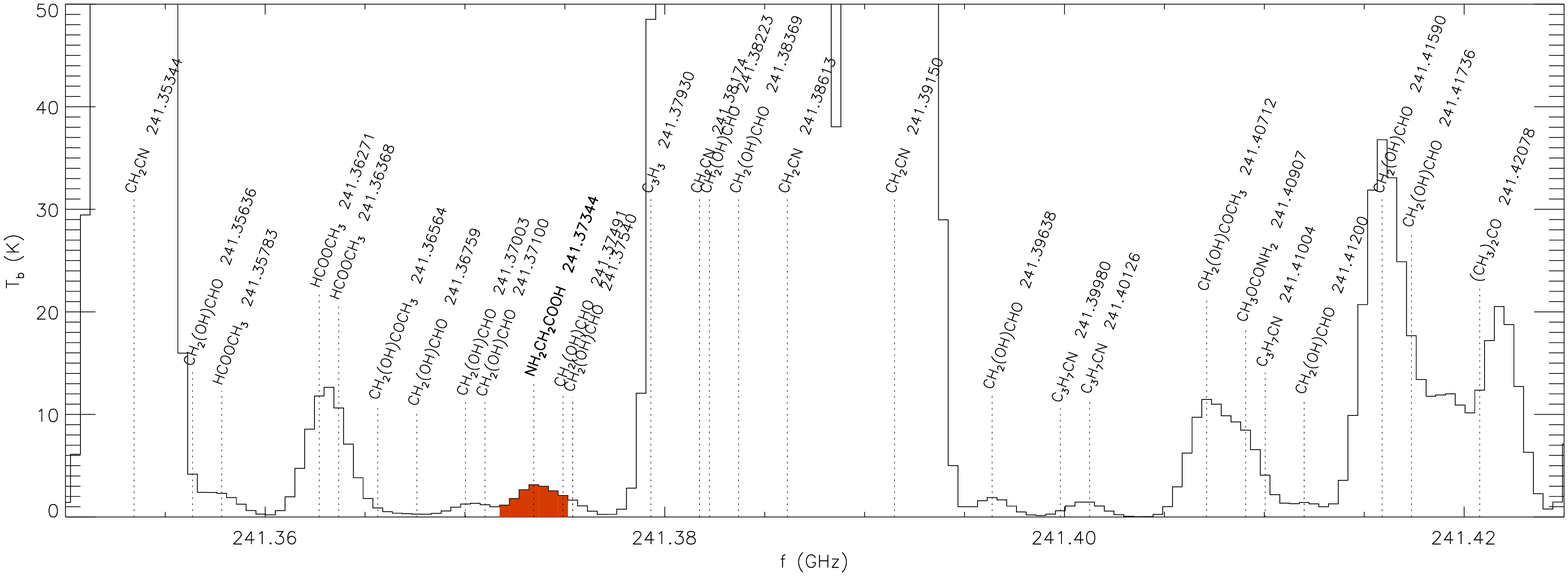}
\includegraphics[width=1.0\textwidth]{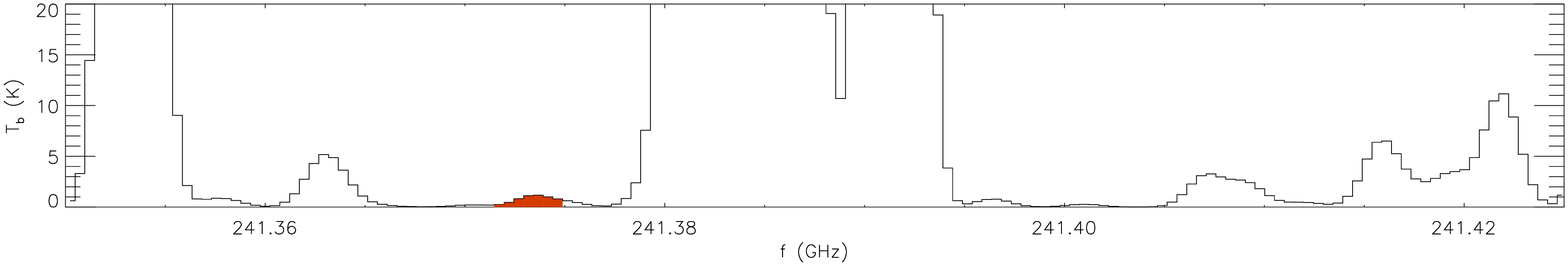}
\includegraphics[width=1.0\textwidth]{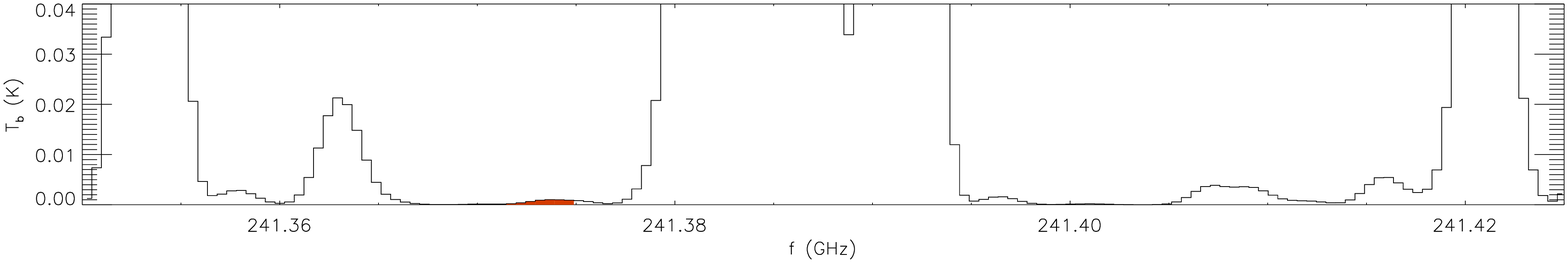}
\caption{\label{fig-specs} Simulated emission spectra for the hot-core source NGC 6334 IRS1 at 241.350 -- 241.425 GHz. {\bf Upper panel}: Local, unconvolved emission for the on-source position (i.e. assuming pencil beam/infinite spatial resolution). {\bf Middle panel}: Convolved emission from entire source, assuming a beam width encompassing the strongest glycine-emission region and appropriate to the ALMA Telescope (0.4 arcsec). {\bf Lower panel}: Convolved emission from entire source, assuming a beam width appropriate to the James Clerk Maxwell Telescope ($\sim$ 20.3 arcsec); note different temperature scale of this panel. All spectra represent on-source pointings. Glycine emission at 241.373 GHz is highlighted in red.}
\end{figure*}

\begin{figure*}
\includegraphics[width=0.475\textwidth]{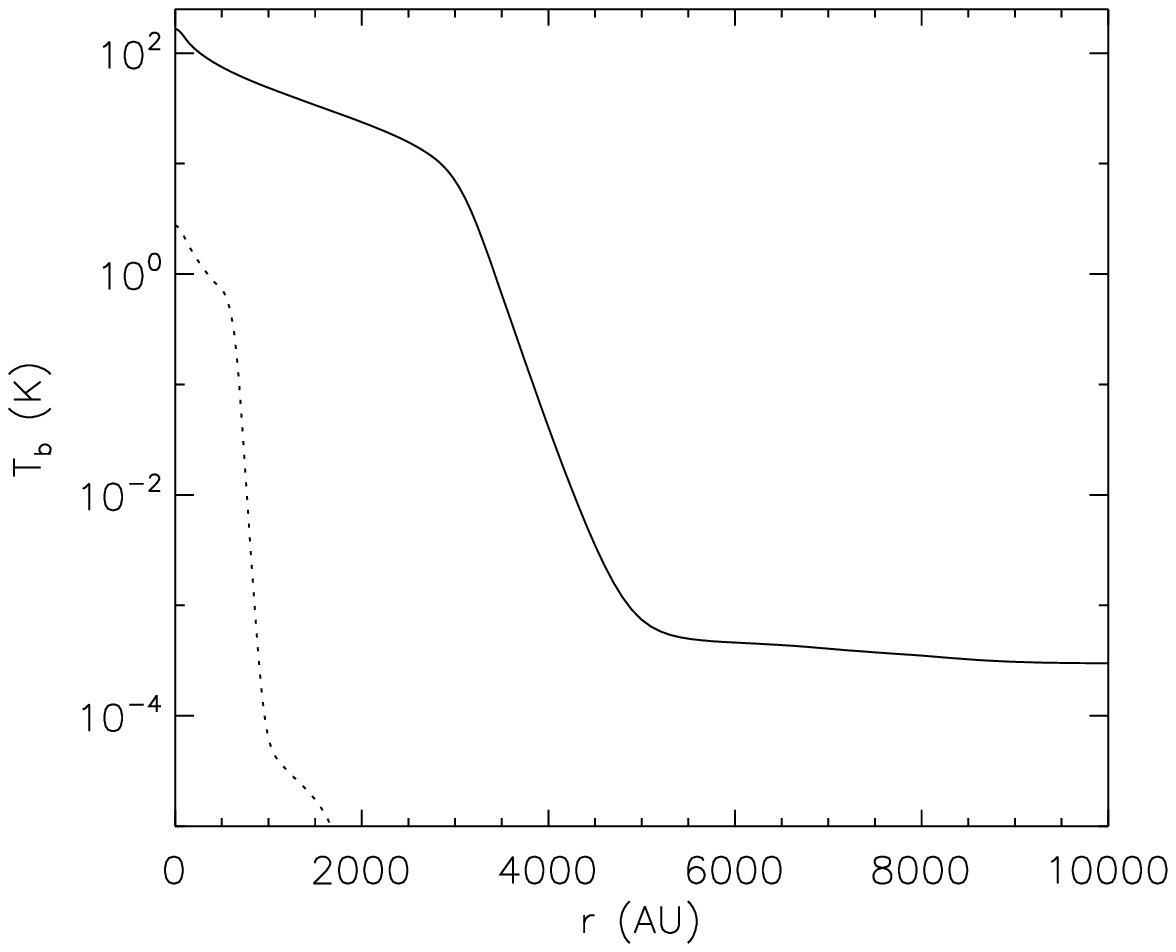}
\includegraphics[width=0.475\textwidth]{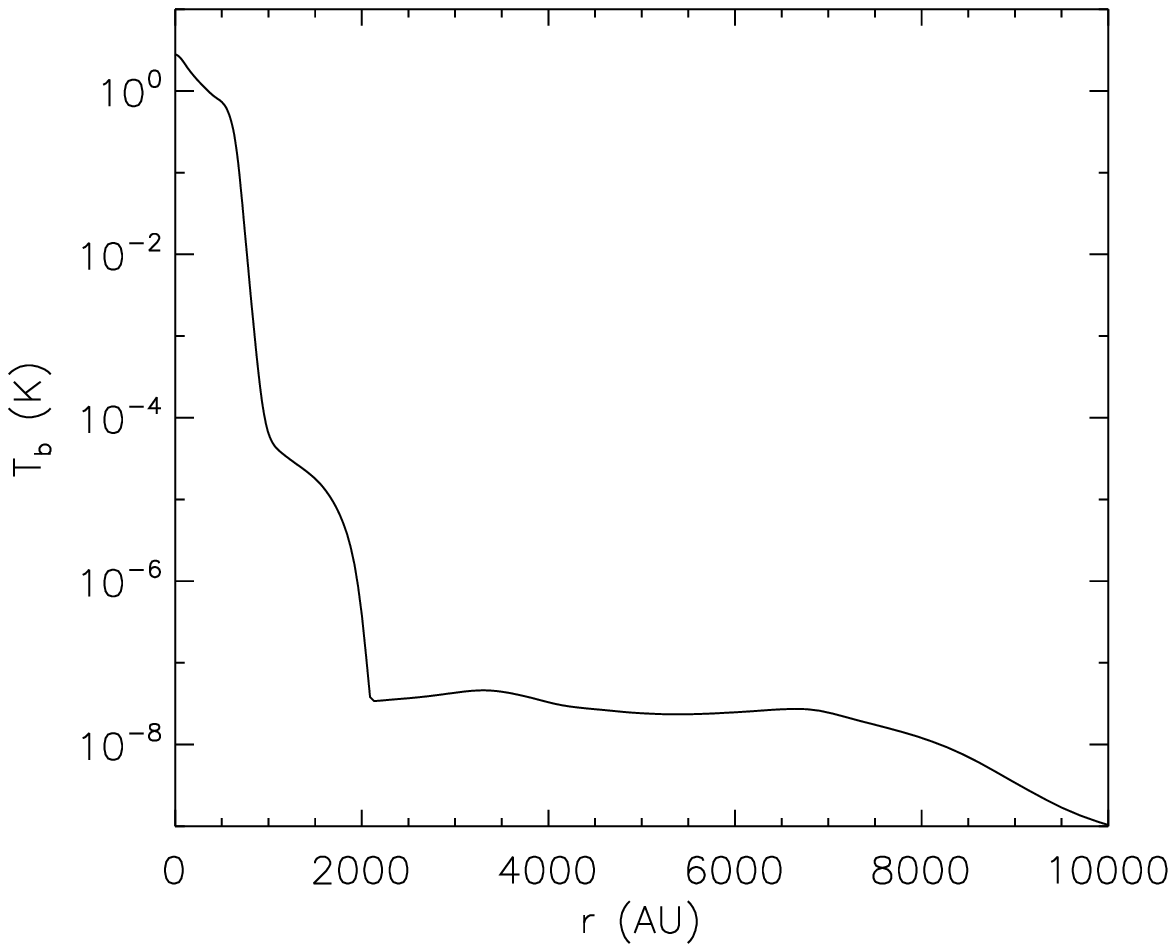}
\caption{\label{fig-profile} Predicted molecular line-strength profiles for lines of sight toward NGC 6334 IRS 1 at radial offsets $r$, assuming a pencil beam. {\bf Left panel:} solid line shows methyl formate emission at 221.979 GHz (dotted line shows glycine emission of right panel, for direct comparison); {\bf Right panel:} glycine emission at 218.105 GHz. Line intensity is expressed as brightness temperature.}
\end{figure*}

\begin{figure*}
\includegraphics[width=0.475\textwidth]{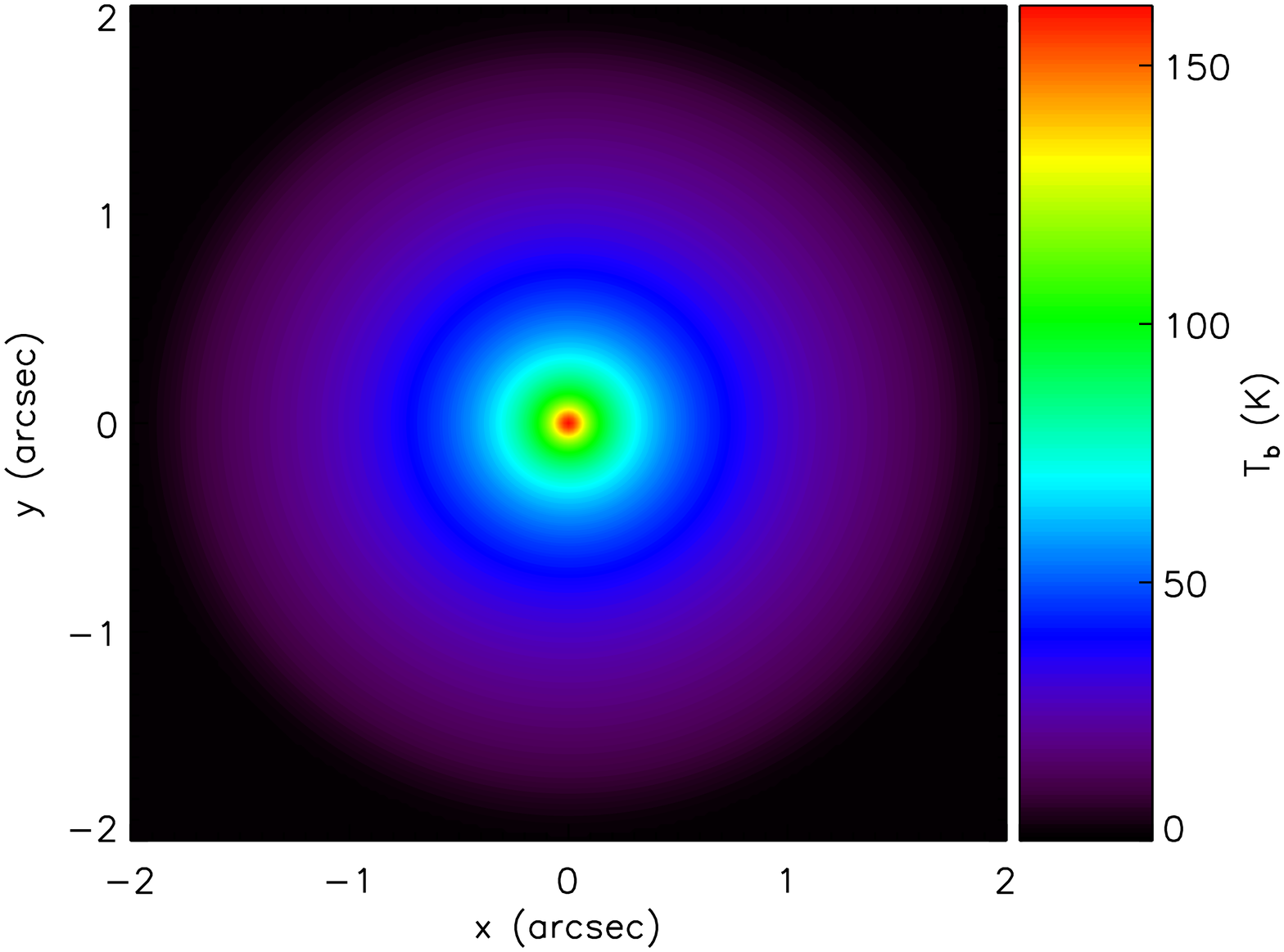}
\includegraphics[width=0.475\textwidth]{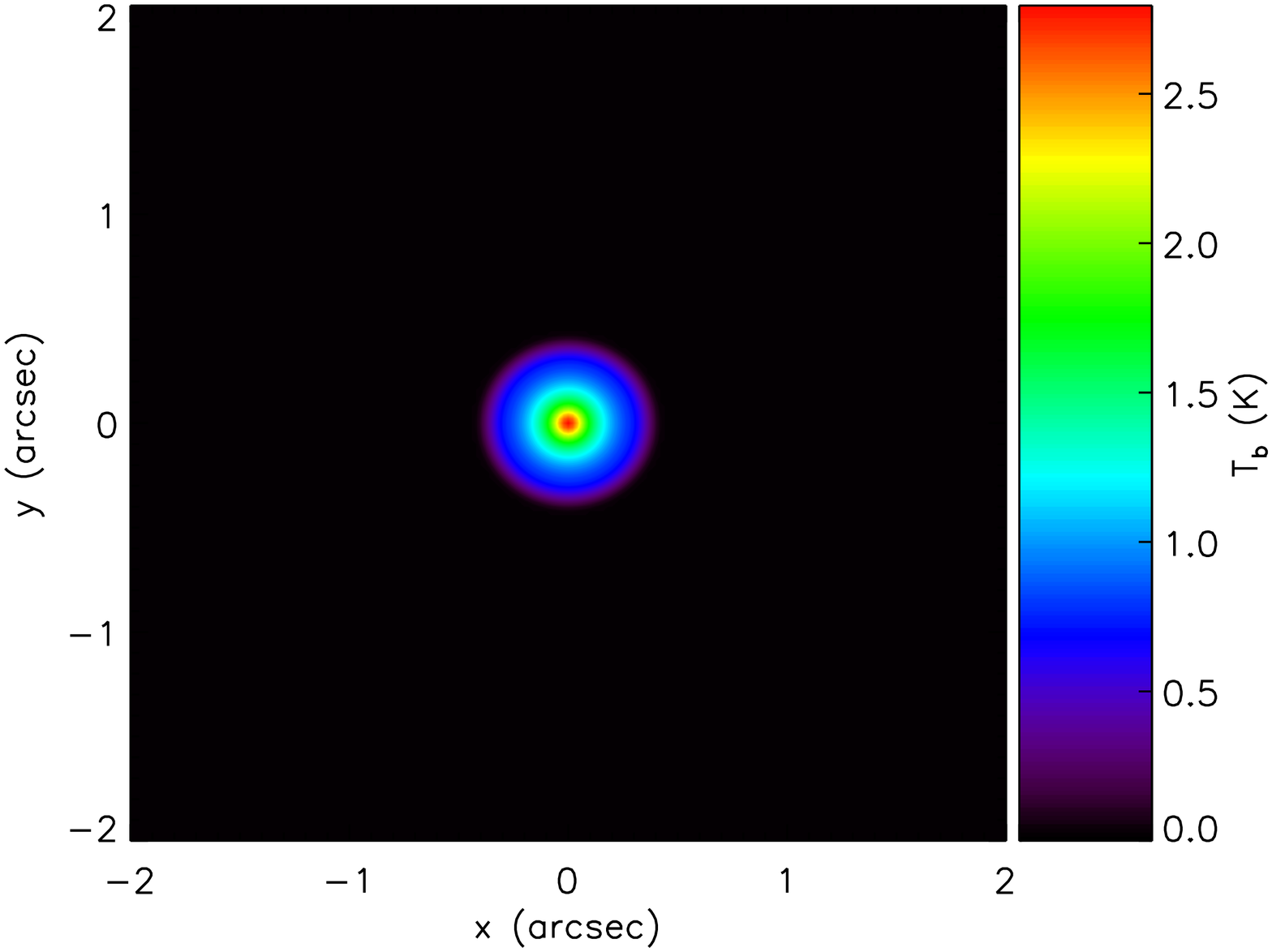}
\caption{\label{fig-map}Predicted molecular line-strength maps for lines of sight toward NGC 6334 IRS 1, assuming a pencil beam (i.e. unconvolved). {\bf Left panel:} methyl formate emission at 221.979 GHz; {\bf Right panel:} glycine emission at 218.105 GHz. Line intensity is expressed as brightness temperature; temperature scales are different in each panel.}
\end{figure*}

\newpage

\begin{deluxetable}{lcccccccccccc}
\tabletypesize{\scriptsize}
\tablecaption{\label{tab-output} Peak fractional abundances and corresponding temperatures for a selection of molecules, for each warm-up model}
\tablewidth{0pt}
\tablecolumns{13}
\tablehead{
 && \multicolumn{3}{c}{Fast} && \multicolumn{3}{c}{Medium} && \multicolumn{3}{c}{Slow} \\
\cline{3-5} \cline{7-9} \cline{11-13} \\
\colhead{Molecule} && \colhead{$n[i]/n_{\mathrm{H}}$} && \colhead{T} && \colhead{$n[i]/n_{\mathrm{H}}$} && \colhead{T} && \colhead{$n[i]/n_{\mathrm{H}}$} && \colhead{T} \\
 && \colhead{} && \colhead{(K)} && \colhead{} && \colhead{(K)} && \colhead{} && \colhead{(K)}
}
\startdata
H$_2$O && 1.6e-4 && 392 && 1.6e-4 && 395 && 1.6e-4 && 398  \\
CO && 9.5e-5 && 400 && 1.0e-4 && 391 && 1.2e-4 && 398  \\
CO$_2$ && 2.9e-5 && 58 && 3.0e-5 && 56 && 3.5e-5 && 55  \\
CH$_4$ && 4.4e-6 && 400 && 4.4e-6 && 398 && 3.1e-6 && 27  \\
HCO$^+$ && 7.1e-10 && 36 && 7.7e-10 && 96 && 8.7e-10 && 56  \\
H$_2$CO && 4.6e-6 && 43 && 3.7e-6 && 42 && 5.5e-7 && 39  \\
CH$_3$OH && 1.1e-5 && 123 && 9.1e-6 && 122 && 3.8e-6 && 121  \\
NH$_3$ && 2.7e-5 && 124 && 2.7e-5 && 122 && 2.7e-5 && 122  \\
\enddata
\end{deluxetable}

\newpage

\begin{deluxetable}{lccccccc}
\tabletypesize{\scriptsize}
\tablecaption{\label{tab-rt-fast} Simulated line intensities toward NGC 6334 IRS1, for a selection of molecules and beamwidths, using {\em fast} warm-up timescale model.}
\tablewidth{0pt}
\tablehead{
\colhead{Molecule} & \colhead{Line} & \colhead{$E_{\mathrm{upper}}$} & \colhead{Peak simulated} & \colhead{Beam} &
\colhead{Convolved} & \colhead{Convolved} & \colhead{Observed}\\
 & \colhead{frequency} & \colhead{} & \colhead{local intensity} & \colhead{width} & \colhead{intensity} & \colhead{integrated intensity} & \colhead{integrated intensity$^a$}\\
 & \colhead{(GHz)} & \colhead{(K)} & \colhead{(K)} & \colhead{(arcsec)} & \colhead{(K)} & \colhead{(K km s$^{-1}$)} &
\colhead{(K km s$^{-1}$)} 
}
\startdata
Methanol               & 220.079  & 96.61 & 113 &        21.9$^b$ &        1.69 &        9.03 & 18  \\
        (CH$_3$OH)     & 230.027  & 39.83 & 118 &        21.0$^b$ &        1.69 &        9.00 & 17  \\
                       & 234.683  & 60.92 & 116 &        20.7$^b$ &        1.76 &        9.40 & 18  \\
                       & 338.125  & 78.08 & 107 &        14.4$^b$ &        4.24 &        22.6 & 56  \\
                       & 338.345  & 70.55 & 106 &        14.4$^b$ &        4.30 &        23.0 & 71  \\
                       & 340.141  & 44.67 & 120 &        14.3$^b$ &        3.34 &        17.8 & 27  \\
\vspace{10pt}          & 356.007  & 295.3 & 107 &        13.9$^b$ &        4.14 &        22.1 & 29  \\

Methyl formate         & 220.190 &        103.1 &        246 &        21.9$^b$ &       0.897 &        4.78 & 6.4  \\
\enddata
\tablecomments{Methanol line data from Xu \& Lovas (1997), using partition function given by the Cologne Database for Molecular Spectroscopy (CDMS); methyl formate data from the JPL linelist based on the dataset of Ilyushin et al. (2009), using partition function given by JPL; glycine line data from Ilyushin et al. (2005), using partition function given by CDMS.}
\tablenotetext{a}{Observations by Bisschop et al. (2007)}
\tablenotetext{b}{Appropriate to JCMT (12 m) observations at this frequency}
\tablenotetext{c}{Appropriate to IRAM\,30\,m observations at this frequency}
\tablenotetext{d}{Approximate value, within ALMA capabilities at this frequency}
\end{deluxetable}
\newpage

\begin{deluxetable}{lccccccc}
\tabletypesize{\scriptsize}
\tablecaption{\label{tab-rt-mid} Simulated line intensities toward NGC 6334 IRS1, for a selection of molecules and beamwidths, using {\em medium} warm-up timescale model.}
\tablewidth{0pt}
\tablehead{
\colhead{Molecule} & \colhead{Line} & \colhead{$E_{\mathrm{upper}}$} & \colhead{Peak simulated} & \colhead{Beam} &
\colhead{Convolved} & \colhead{Convolved} & \colhead{Observed}\\
 & \colhead{frequency} & \colhead{} & \colhead{local intensity} & \colhead{width} & \colhead{intensity} & \colhead{integrated intensity} & \colhead{integrated intensity$^a$}\\
 & \colhead{(GHz)} & \colhead{(K)} & \colhead{(K)} & \colhead{(arcsec)} & \colhead{(K)} & \colhead{(K km s$^{-1}$)} &
\colhead{(K km s$^{-1}$)} 
}
\startdata
Methanol               & 220.079  & 96.61 & 112 &        21.9$^b$ &        1.71 &        9.14 & 18  \\
        (CH$_3$OH)     & 230.027  & 39.83 & 118 &        21.0$^b$ &        1.72 &        9.19 & 17  \\
                       & 234.683  & 60.92 & 116 &        20.7$^b$ &        1.80 &        9.59 & 18  \\
                       & 338.125  & 78.08 & 106 &        14.4$^b$ &        4.17 &        22.2 & 56  \\
                       & 338.345  & 70.55 & 106 &        14.4$^b$ &        4.21 &        22.5 & 71  \\
                       & 340.141  & 44.67 & 122 &        14.3$^b$ &        3.44 &        18.4 & 27  \\
\vspace{10pt}          & 356.007  & 295.3 & 106 &        13.9$^b$ &        4.25 &        22.7 & 29  \\

Methyl formate         & 220.190 & 103.1 & 133 &        21.9$^b$ &       0.458 &        2.44 & 6.4  \\
\enddata
\tablecomments{Methanol line data from Xu \& Lovas (1997), using partition function given by the Cologne Database for Molecular Spectroscopy (CDMS); methyl formate data from the JPL linelist based on the dataset of Ilyushin et al. (2009), using partition function given by JPL; glycine line data from Ilyushin et al. (2005), using partition function given by CDMS.}
\tablenotetext{a}{Observations by Bisschop et al. (2007)}
\tablenotetext{b}{Appropriate to JCMT (12 m) observations at this frequency}
\tablenotetext{c}{Appropriate to IRAM\,30\,m observations at this frequency}
\tablenotetext{d}{Approximate value, within ALMA capabilities at this frequency}

\end{deluxetable}
\newpage

\begin{deluxetable}{lccccccc}
\tabletypesize{\scriptsize}
\tablecaption{\label{tab-rt-slow} Simulated line intensities toward NGC 6334 IRS1, for a selection of molecules and beamwidths, using {\em slow} warm-up timescale model.}
\tablewidth{0pt}
\tablehead{
\colhead{Molecule} & \colhead{Line} & \colhead{$E_{\mathrm{upper}}$} & \colhead{Peak simulated} & \colhead{Beam} &
\colhead{Convolved} & \colhead{Convolved} & \colhead{Observed}\\
 & \colhead{frequency} & \colhead{} & \colhead{local intensity} & \colhead{width} & \colhead{intensity} & \colhead{integrated intensity} & \colhead{integrated intensity$^a$}\\
 & \colhead{(GHz)} & \colhead{(K)} & \colhead{(K)} & \colhead{(arcsec)} & \colhead{(K)} & \colhead{(K km s$^{-1}$)} &
\colhead{(K km s$^{-1}$)} 
}
\startdata
Methanol               & 220.079  & 96.61 & 117 &        21.9$^b$ &        1.65 &        8.82 & 18  \\
        (CH$_3$OH)     & 230.027  & 39.83 & 127 &        21.0$^b$ &        1.63 &        8.67 & 17  \\
                       & 234.683  & 60.92 & 126 &        20.7$^b$ &        1.73 &        9.25 & 18  \\
                       & 338.125  & 78.08 & 107 &        14.4$^b$ &        3.97 &        21.2 & 56  \\
                       & 338.345  & 70.55 & 107 &        14.4$^b$ &        4.00 &        21.3 & 71  \\
                       & 340.141  & 44.67 & 116 &        14.3$^b$ &        2.99 &        15.9 & 27  \\
\vspace{10pt}          & 356.007  & 295.3 & 108 &        13.9$^b$ &        4.16 &        22.2 & 29  \\

Methyl formate         & 220.190 & 103.1 & 0.773 &       21.9$^b$ &      0.0152 &     0.0813 & 6.4  \\
\enddata
\tablecomments{Methanol line data from Xu \& Lovas (1997), using partition function given by the Cologne Database for Molecular Spectroscopy (CDMS); methyl formate data from the JPL linelist based on the dataset of Ilyushin et al. (2009), using partition function given by JPL; glycine line data from Ilyushin et al. (2005), using partition function given by CDMS.}
\tablenotetext{a}{Observations by Bisschop et al. (2007)}
\tablenotetext{b}{Appropriate to JCMT (12 m) observations at this frequency}
\tablenotetext{c}{Appropriate to IRAM\,30\,m observations at this frequency}
\tablenotetext{d}{Approximate value, within ALMA capabilities at this frequency}

\end{deluxetable}
\newpage

\begin{deluxetable}{lcccccc}
\tabletypesize{\small}
\tablecaption{\label{tab-nh2oh-1} Simulated line intensities for hydroxylamine (NH$_2$OH) toward W3 IRS5, using standard {\em medium} warm-up model}
\tablewidth{0pt}
\tablehead{
\colhead{Molecule} & \colhead{Line} & \colhead{$E_{\mathrm{upper}}$} & \colhead{Peak simulated} & \colhead{Beam} &
\colhead{Convolved} & \colhead{Convolved} \\
 & \colhead{frequency} & \colhead{} & \colhead{local intensity} & \colhead{width$^a$} & \colhead{intensity} & \colhead{integrated intensity}\\
 & \colhead{(GHz)} & \colhead{(K)} & \colhead{(K)} & \colhead{(arcsec)} & \colhead{(K)} & \colhead{(K km s$^{-1}$)} 
}
\startdata
Hydroxylamine & 151.102 &  46.33 &        0.0354 &        41.5 &   4.37e-5 &    4.20e-4   \\
(NH$_2$OH)    & 151.118 &  14.51 &        0.0391 &        41.5 &   5.01e-5 &    4.81e-4   \\
                     &  164.341 &  116.7 &      8.05e-4 &        38.2 &   1.08e-6 &   1.03e-5   \\
                     &  164.627 &  94.92 &      8.10e-4 &        38.1 &   1.12e-6 &   1.07e-5   \\
                     &  164.883 &  75.60 &      7.93e-4 &        38.0 &   1.13e-6 &   1.08e-5  
\enddata
\tablecomments{* Line data and partition function obtained from the Cologne Database for Molecular Spectroscopy (CDMS).}
\tablenotetext{a}{Appropriate to NRAO 12 m observations at this frequency.}

\end{deluxetable}

\begin{deluxetable}{lcccccc}
\tabletypesize{\small}
\tablecaption{\label{tab-nh2oh-2} Simulated line intensities for hydroxylamine (NH$_2$OH) toward W3 IRS5, using {\em medium} warm-up model with the gas-phase glycine-formation mechanism and high initial NH$_2$OH ice-mantle abundance discussed in Sec. 4.1.}
\tablewidth{0pt}
\tablehead{
\colhead{Molecule} & \colhead{Line} & \colhead{$E_{\mathrm{upper}}$} & \colhead{Peak simulated} & \colhead{Beam} &
\colhead{Convolved} & \colhead{Convolved} \\
 & \colhead{frequency} & \colhead{} & \colhead{local intensity} & \colhead{width$^a$} & \colhead{intensity} & \colhead{integrated intensity}\\
 & \colhead{(GHz)} & \colhead{(K)} & \colhead{(K)} & \colhead{(arcsec)} & \colhead{(K)} & \colhead{(K km s$^{-1}$)} 
}
\startdata
Hydroxylamine & 151.102 &  46.33 &        34.7 &        41.5 &   0.0424 &    0.407   \\
(NH$_2$OH)    & 151.118 &  14.51 &        37.7 &        41.5 &   0.0494 &    0.475   \\
                     &  164.341 &  116.7 &      0.883 &        38.2 &  0.00111 &   0.0107   \\
                     &  164.627 &  94.92 &      0.885 &        38.1 &   0.00115 &   0.0111   \\
                     &  164.883 &  75.60 &      0.865 &        38.0 &   0.00117 &   0.0112  
\enddata
\tablecomments{* Line data and partition function obtained from the Cologne Database for Molecular Spectroscopy (CDMS).}
\tablenotetext{a}{Appropriate to NRAO 12 m observations at this frequency.}

\end{deluxetable}

\begin{figure}

\end{figure}


\end{document}